\documentclass[pdflatex,sn-mathphys-num]{sn-jnl}

\usepackage{graphicx}%
\usepackage{multirow}%
\usepackage[font=footnotesize]{caption}
\usepackage{amsmath,amssymb,amsfonts, algorithm}
\usepackage{amsthm}%
\usepackage{mathrsfs}%
\usepackage[title]{appendix}%
\usepackage{xcolor}%
\usepackage{textcomp}%
\usepackage{manyfoot}%
\usepackage{booktabs}%
\usepackage{algorithmicx}%
\usepackage{algpseudocode}%
\usepackage{listings}%

\theoremstyle{thmstyleone}%
\theoremstyle{thmstyletwo}%

\theoremstyle{thmstylethree}%

\usepackage{graphicx}
\usepackage{textcomp}
\usepackage{xcolor}
\usepackage{hyperref}
\usepackage{booktabs}
\usepackage{siunitx}
\usepackage{subcaption}
\usepackage{soul}
\usepackage{url}
\usepackage{colortbl}
\usepackage{multirow}
\usepackage[utf8]{inputenc}
\usepackage{pifont}
\usepackage{tikz}
\usepackage{newunicodechar}
\newunicodechar{✓}{\ding{51}}
\newunicodechar{✗}{\ding{55}}

\begin{document}

\title[Article Title]{Fully Homomorphic Encryption on Llama 3 model for privacy preserving LLM inference}


\author*[1]{\fnm{Anes} \sur{Abdennebi}}\email{anes.abdennebi.1@ens.etsmtl.ca}

\author[1]{\fnm{Nadjia} \sur{Kara}}\email{Nadjia.Kara@etsmtl.ca}
\equalcont{These authors contributed to this work by supervising the methodology and revising the manuscript.}

\author[1]{\fnm{Laaziz} \sur{Lahlou}}\email{laaziz.lahlou@etsmtl.ca}
\equalcont{These authors contributed to this work by supervising the methodology and revising the manuscript.}

\affil*[1]{\orgdiv{Software and IT Engineering Department}, \orgname{École de Technologie Supérieure}, \orgaddress{\city{Montreal}, \state{Quebec}, \country{Canada}}}




\abstract{The applications of Generative Artificial Intelligence (GenAI) and their intersections with data-driven fields, such as healthcare, finance, transportation, and information security, have led to significant improvements in service efficiency and low latency. However, this synergy raises serious concerns regarding the security of large language models (LLMs) and their potential impact on the privacy of companies and users' data. Many technology companies that incorporate LLMs in their services with a certain level of command and control bear a risk of data exposure and secret divulgence caused by insecure LLM pipelines, making them vulnerable to multiple attacks such as data poisoning, prompt injection, and model theft. 
	Although several security techniques (input/output sanitization, decentralized learning, access control management, and encryption) were implemented to reduce this risk, there is still an imminent risk of quantum computing attacks, which are expected to break existing encryption algorithms, hence, retrieving secret keys, encrypted sensitive data, and decrypting encrypted models. 
	In this extensive work, we integrate the Post-Quantum Cryptography (PQC) based Lattice-based Homomorphic Encryption (HE) main functions in the LLM’s inference pipeline to secure some of its layers against data privacy attacks. We modify the inference pipeline of the transformer architecture for the \textsc{LLAMA-3} model while injecting the main homomorphic encryption operations provided by the \texttt{concrete\--ml} library. We demonstrate high text generation accuracies (up to $98\%$) with reasonable latencies ($237$ ms) on an i9 CPU, reaching up to $80$ tokens per second, which proves the feasibility and validity of our work while running a FHE-secured \textsc{LLAMA-3} inference model. Further experiments and analysis are discussed to justify models' text generation latencies and behaviours.}


\maketitle

\bmhead{Acknowledgements}
This work was supported by the Department of National Defence of Canada under the Mobilizing Insights in Defence and Security (MINDS - Targeted Engagement Grant) program, Grant No 24-2-62.

\section{Introduction}
Large Language Models and their building block, the transformer architecture, are the current cutting-edge tools that boosted different Natural Language Processing (NLP) tasks, such as text generation, translation, summarization, sequence classification, and other evolving textual tasks. The transformers, coming with different variants (encoder-decoder, encoder only, or decoder only), depend on the self-attention mechanism implemented in layers forming the majority of the architecture. This mechanism focuses on generating the next words (or tokens) while assigning weights and degrees of relevance to contextual words fed to the model. It is the core engine of any LLM that ensures accurate and pertinent text that makes sense within the scope of the desired task. As LLMs offer their linguistic utility in different tasks, their applicability fits perfectly with the needs of several fields, in which a surge in LLM usage is witnessed in many markets. 

The proliferation of Large Language Model-dependent applications in various fields, such as market trading, finance, healthcare, and transportation, for applications ranging from sentiment analysis, market strategy simulation, illness symptoms analysis, insight extraction, traffic patterns, system logs analysis, and a plethora of other applications. The employment of LLMs takes different forms, it can be the core of an application as a full software based on a fine-tuned LLM (Software-as-a-Service), a partial code running its data on a model, or the most prevalent form of usage as chatbots.
Regardless of the application scheme in which the large language models are used, these apps employ and communicate data through their inner LLM architectures and inference pipelines. Moreover, these data would most likely incorporate sensitive information about individuals, corporations, companies, and government facilities. The processing of such data in plain (not encrypted) form within LLMs poses a great risk of breaching users' personal and sensitive data. For instance, running a query on a locally deployed chatbot specialized for providing a preliminary analysis of a patient's condition, while containing their disease information, blood tests results, magnetic resonance imaging (MRI), and other test data, would form a significant threat to the patients' data confidentiality and integrity. The examples extend to other application scenarios where sensitive data in plain text with LLMs is recklessly used.
This forms a significant threat in the near future, especially since almost $50\%$ of applications are depending on LLMs in $2025$ \cite{bib1} .

Attackers and malicious actors seeking the exploitation of such apps (that use LLMs in their backend) tend to use sophisticated techniques to successfully deceive the models into generating the desired outputs in scenarios where their main goal is to extract, distort, or recover sensitive information. The leveraged techniques include prompt injection, jailbreaks, model and data poisoning, data extraction, and supply chain attacks through unauthorized access to the system or API \cite{singh2023exploiting,mozes2023use,yao2024survey,choquet2024exploiting}.  Attackers may use one, a few, or a combination of these techniques to achieve their goal \cite{kang2024exploiting}.

The attacking techniques are not limited to plain queries; they also comprise different vulnerabilities within LLMs. There are multiple ways of manipulating the transformer architecture of LLMs in order to generate malicious content or expose important information about the data used in training the pre-trained model, or even divulge the instructions given to the generative model itself \cite{zou2023universal,nasr2023scalable,greshake2023not,perez2022ignore}. All the previously mentioned risks give attackers an advantage in maximizing the exploitation of their target models. Furthermore, another attack category that falls within the data privacy circle is the Private Information Extraction (PIE) attack, where, in such a scenario, malicious actors can apply techniques to dig deep into LLMs to retrieve information related to any users who had previous interactions (query-answer) with the model. Information including names, birth dates, social insurance numbers, credit cards numbers, passwords, companies confidential records (especially when employees use generative AI chatbots to help them speedup their data analysis, reports generation, or any other time-consuming task), and even extract links between users/clients and their affiliations and/or enterprises \cite{kpmgImpactsArtificial} potentially leading to a high risk of divulging critical and confidential data.      

The risk of violating data privacy in LLMs and their variants, especially when they manifest into existing threats posed by cybercriminals, is severe. It requires stringent techniques to tighten the confidentiality of such data when shared with these vulnerable models. The work in \cite{huang2024genai} offers a comprehensive framework based on an application security practice (OWASP Top 10 for LLMs), suggesting how to secure LLMs against a wide set of attacks, including \textit{prompt injection}, \textit{insecure output}, \textit{model theft}, and \textit{data poisoning}. However, the work omits the overhead added when applying these theoretically proposed mitigation actions. Moreover, there is no real application that is proven to be effective against one of the mentioned LLM attacks.
Another paper by Deng et al. \cite{deng2023masterkey} shows the possibility of attacking LLM-based chatbots such as GPT-3.5, GPT-4, and Gemini (previously known as Bard). It bypasses their security measures through jailbreaks and time-based analysis to reverse engineer their defensive mechanisms. The curated attacks by authors' framework (\textsc{MASTERKEY}) achieved an over $21\%$ success rate, leading the chatbots service providers and research community to develop new and enhance existing security strategies and mechanisms based on the outcomes of this work.  
Many other works are focusing on attacking LLM-based chatbots or tools using jailbreak techniques, experimenting mainly with GPT-3.5, GPT-4, Gemini, Claude, Grok, and other models with natural language processing capabilities, while providing insights on LLMs security reinforcement spots \cite{nguyen2025penetration,10992337,yi2024jailbreak,peng2024jailbreaking,jiang2024wildteaming}.

A promising direction to strengthen the model and data's privacy is to use encryption algorithms to encrypt the data at the different levels and stages of the models training and testing, which is the case with many machine learning (ML) and deep learning (DL) models such as RandomForest (RF), Decision Trees (DT), XGBoost (XGB), and other deep neural network models.
Using secure encryption algorithms on the data, the model, or both is one solution to preserve data privacy and ensure its sanity within its legitimate environments, especially when used in training or querying large language models. Several prior works have implemented encryption on ML and DL models, securing the training process, the inference pipeline, and the data fed to the model \cite{chen2021privacy,zuo2021sealing,huang2022privacy,balaban2025privacy,cong2022sortinghat,akhavan2023level,bost2014machine,graepel2012ml,gonzalez2018supervised,shen2022machine,frimpong2024guardml}. However, when it comes to LLMs to apply homomorphic encryption, the process is sensitive and complex due to the task complexity and the anticipated computational overhead imposed by expensive HE operations. Therefore, attempting to secure an LLM internally within its attention mechanism should be approached carefully, and only the concerned components should be secured to avoid unnecessary overhead while maintaining the desired security level.     

In this work, we focus on securing the LLM's inference stage since integrating computationally expensive cryptographic operations into the training process would not be practical and excessively time-consuming. We leverage one of the post-quantum algorithms in fully homomorphic encryption (FHE) to secure the inference process of a specific large language model, \textsc{Llama-3} \cite{Llama-3-8B}. We integrate a specific fully homomorphic encryption scheme offered within the Zama's library\footnote{An open-source cryptography company pioneering Fully Homomorphic Encryption for confidential applications including AI.}, \texttt{concrete\--ml}\cite{ConcreteML}, which is suitable for the transformers architecture of \textsc{Llama-3} to run the inference process in fully encrypted mode, hence, secure the data and inference pipeline for selected model's layers. Another motivation for this work is to assess the hardware requirements and costs for such a computational expensive integration.

The rest of the paper is organized as follows: We provide a brief mathematical background about FHE and its main operations, citing the main homomorphic encryption schemes in Section~\ref{sec:pre}, and an overview of \texttt{LLAMA-3} model's architecture. Section~\ref{sec:method} dissects the transformer's architecture, showing our full FHE integration methodology within its inference components. Section~\ref{sec:exp} presents the experimental setup and the results of running our homomorphically encrypted model. Finally, we discuss the use cases of our PQC-secured model in Section~\ref{sec:qllama_llm}, then conclude the work in Section~\ref{sec:conc} to draw a comprehensive conclusion about this work.

\section{Preliminaries}\label{sec:pre}
Fully Homomorphic Encryption is a cryptographic paradigm that enables arbitrary computations on encrypted data without revealing the underlying plaintexts. some FHE schemes rely on the hardness of the \emph{Ring Learning With Errors} (RLWE) problem \cite{lyubashevsky2010ideal,lyubashevsky2013ideal}.  
In RLWE, for a secret polynomial $s \in \mathbb{Z}_q[X]/(X^n+1)$ and a small noise polynomial $e$, an adversary is given samples $(a_i, b_i = a_i s + e_i) \in (\mathbb{Z}_q[X]/(X^n+1))^2$ and must recover $s$.  
The computational intractability of this problem provides the security foundation of modern lattice-based FHE: even if an adversary observes many ciphertexts and intermediate homomorphic computations, recovering plaintexts or secret keys is computationally infeasible, as both $a_i$ and  $b_i = a_i s + e_i)$ look absolutely random, which makes it hard to distinguish between them, hence, recover the secret $s$.  

Practical FHE implementations, such as those in the \texttt{concrete-ml} library, leverage RLWE-based schemes to perform linear transformations (matrix-vector multiplications) and certain nonlinear operations (e.g., approximate ReLU) entirely in the encrypted domain. 
Techniques such as \emph{ciphertext packing}, \emph{relinearization}, and \emph{bootstrapping} enable parallelism and control noise growth, making it feasible to evaluate large neural networks, including LLMs, without exposing sensitive input embeddings or intermediate weights or activations.

For the rest of the paper, we refer to homomorphic encryption as HE for simplicity, and when providing definitions or mathematical representations, our work depends on a fully homomorphic encryption implementation, referred to as FHE, thereby, both pertain to the same implementation.  


\subsection{Mathematical Definition}\label{subsec:math}

A homomorphic encryption scheme is a tuple of probabilistic polynomial-time algorithms:

\[
\mathsf{HE} = (\mathsf{KeyGen}, \mathsf{Enc}, \mathsf{Dec}, \mathsf{Eval})
\]

with the following functionalities:

\begin{itemize}
	\item $\mathsf{KeyGen}(1^\lambda) \rightarrow (\mathsf{pk}, \mathsf{ek}, \mathsf{sk})$: Given a security parameter $\lambda$, outputs a public key $\mathsf{pk}$ and secret key $\mathsf{sk}$, and an evaluation key $\mathsf{ek}$. The latter is used while performing operations on encrypted data, it can be distributed in plain.
	\item $\mathsf{Enc}_{\mathsf{pk}}(m) \rightarrow c$: Given a message or a plain input $m \in \mathcal{M}$, returns a ciphertext $c \in \mathcal{C}$.
	\item $\mathsf{Dec}_{\mathsf{sk}}(c) \rightarrow m$: Given a ciphertext $c$ and a private key $\mathsf{sk}$, returns the plaintext message $m$.
	\item $\mathsf{Eval}_{\mathsf{ek}}(f, c_1, \ldots, c_n) \rightarrow c_f$: Given a function $f$ and ciphertexts $c_1, \ldots, c_n$ corresponding to plaintexts $m_1, \ldots, m_n$, outputs a ciphertext $c_f$ such that:
	\[
	\mathsf{Dec}_{\mathsf{sk}}(c_f) = f(m_1, \ldots, m_n)
	\]
\end{itemize}

Homomorphic encryption bears several properties regarding the operations it supports (addition and multiplication):

Let $\circ$ be a binary operation on plaintext space $\mathcal{M}$ and $\ast$ the corresponding operation on ciphertexts. A scheme is said to be \textit{homomorphic} with respect to $\circ$ if:

\[
\mathsf{Dec}_{\mathsf{sk}} \left( \mathsf{Enc}_{\mathsf{pk}}(m_1) \ast \mathsf{Enc}_{\mathsf{pk}}(m_2) \right) = m_1 \circ m_2
\]

In FHE constructions, noise is an essential security parameter introduced during encryption to mask the underlying plaintext from algebraic recovery. A ciphertext is typically represented as an element $c$ that satisfies a relation such as $c \equiv s \cdot a + e + m \pmod q$, where $m$ is the message and $e$ is a small error term. As homomorphic operations are performed, this noise scales according to the complexity of the circuit: for addition, the resulting noise of adding two ciphertexts ($c_{1}$ and $c_{2}$) is roughly the sum of their respective error values ($e_{add} \approx e_1 + e_2$), whereas for multiplication, the noise grows at a much higher rate, becoming approximately the product of the existing errors ($e_{mult} \approx e_1 \cdot e_2$). If the cumulative magnitude of $e$ exceeds a predetermined threshold dictated by the modulus $q$, the noise overlaps with the message bits, leading to decryption failures and rendering the data unrecoverable. In this work, we refer to noise by either $e$ or $\eta$.

\subsubsection*{\textbf{Boolean circuits}} Since encryption operations include additions and multiplications, the series of operations can represented in a boolean circuit of \textbf{\texttt{AND}}s and \textbf{\texttt{XOR}}s within the plaintext space $\mathcal{M \in \mathbb{Z_{\mathsf{2}}}}$ represented in bits and $\mod 2$. Having the homomorphic operations represented in a circuit of universal \textbf{\texttt{NAND}} gates (universal since they can represent any logical function), allows estimating a noise budget (how many operations can be performed homomorphically before exceeding the ciphertext correctness threshold on decryption).
\subsubsection*{\textbf{Arithmetic circuits}} is also an option to represent the operands in integer data types and perform additions and multiplications homomorphically in \texttt{modulo} $p$ where $p$ is a prime number greater than $2$.

Given the main properties of HE, and their potential circuit representation, it is substantial to distinguish several types of homomorphic encryption approaches based on the operations they support, the multiplicative depth, and their estimated increasing noise:

\begin{enumerate}
	\item \textbf{Partially Homomorphic Encryption (PHE)}: Supports only one operation (either addition or multiplication), such as Rivest-Shamir-Adleman-RSA (multiplicative) and Paillier (additive).
	
	\item \textbf{Somewhat Homomorphic Encryption (SHE)}: Supports both operations but only for a limited depth (limited complexity of the $f$ function).
	
	\item \textbf{Leveled Fully Homomorphic Encryption (Leveled FHE)}: Supports all operations for functions of bounded depth without noise refreshing.
	
	\item \textbf{Fully Homomorphic Encryption (FHE)}: Supports arbitrary depth computations, often achieved by incorporating a \textit{bootstrapping} technique to reduce noise in ciphertexts that increases exponentially after several operations. This operation refreshes the ciphertexts so their decryption would not be corrupted due to uncontrolled noise levels.
\end{enumerate}

\subsection{Homomorphic Encryption Schemes}\label{sec:fhesc}

HE has different schemes applicable to various applications in accordance with the type of operations required for each application. Some perform operations on exact integers $\mathbb{Z}$, while other applications demand operations on approximate real numbers $\mathbb{R}$, such as in our study, since transformers are built on neural networks performing multiplication and addition operations in their network units incorporated within the attention mechanism, in addition to activation, softmax, and other non-linear functions.

Homomorphic Encryption, as a powerful tool for privacy-preserving applications, remained theoretical until Craig Gentry introduced the first FHE scheme in 2009~\cite{gentry2009}, which enabled the evaluation of arbitrary Boolean circuits on encrypted data. 
This initial scheme allowed operations such as negation, multiplication, and addition to be performed on ciphertexts without revealing the original plaintexts. Since then, the field has witnessed extensive research and the emergence of further FHE schemes based on diverse hardness assumptions \cite{gentry2009, van2010fully, brakerski2014leveled, brakerski2012fully, lopez2012fly, gentry2013homomorphic}. These schemes share critical limitations, including the noise increase due to expensive homomorphic operations within ciphertexts. While addition and multiplication in the binary field $F_{2}$ form the basis for Boolean circuit evaluation, they amplify the noise, especially multiplication. This growing noise eventually increases the chances of having incorrect decryption unless avoided by an intermediate operation, bootstrapping. 
Bootstrapping is a costly procedure for restoring the ciphertext's noise to acceptable levels and preserving ciphertext usability. Due to its computational intensity, only a few FHE schemes have been implemented in practice \cite{gentry2011implementing, cheon2013batch}, and those suffer from unsatisfactory performance. Furthermore, with each noise mitigation operation comes an additional memory burden as more storage for ciphertext polynomials and/or new matrices for matrix-matrix and vector-matrix multiplication are required. This implies that either CPUs or whatever hardware accelerators (GPUs or FPGAs) are used to run the FHE scheme should have enough vRAM capacity.

Recognizing that many real-world applications do not require arbitrary circuit evaluation, researchers explored alternative designs that forgo bootstrapping. These designs target circuits with low and known multiplicative depth, reducing computational overhead \cite{naehrig2011can,bos2013improved,gentry2012homomorphic}. However, schemes such as \cite{gentry2011implementing,coron2012public,cheon2013batch} face exponential noise growth with circuit depth, significantly limiting their scalability. Brakerski, Gentry, and Vaikuntanathan proposed leveled homomorphic encryption schemes \cite{brakerski2014leveled}, where the noise increases linearly rather than exponentially with the multiplicative depth. Their BGV scheme employs a modulus switching technique and supports plaintexts as polynomials over a ring modulo an integer, enabling batching and larger plaintext spaces. The BGV scheme demonstrated practical use in homomorphic AES evaluation \cite{gentry2012homomorphic}, albeit with high resource demands. Specifically, evaluating a depth-d circuit requires storing d variants of an evaluation key, necessitating up to 256 GB of RAM for AES.

Brakerski introduced the concept of scale-invariance \cite{brakerski2012fully} at Crypto 2012 in response to the memory and efficiency concerns of modulus switching. 
Scale-invariant schemes maintain a constant modulus throughout the computation, needing only a single evaluation key copy. This concept was integrated into BGV by Fan and Vercauteren \cite{fan2012somewhat} and into the LTV scheme \cite{lopez2012fly} by Bos, Lauter, Loftus, and Naehrig \cite{bos2013improved}, resulting in the FV and YASHE \cite{bos2013improved} schemes. While FV remains largely unimplemented, aside from a basic prototype in \cite{graepel2012ml}, YASHE became the first operational scale-invariant leveled homomorphic encryption scheme. It achieved promising performance for shallow circuits on personal computers, though it supported only circuits of multiplicative depth up to two due to small modulus constraints. 

The \texttt{concrete\--ml} library utilizes a specialized refresh mechanism called \textit{Programmable Bootstrapping (PBS)} that simultaneously resets the noise level while evaluating non-linear functions through a lookup-table approach. This process takes a noisy ciphertext representing a message $m$ and outputs a refreshed ciphertext representing $f(m)$, where $f$ is an arbitrary function such as a sigmoid or a rectified linear unit (ReLU). Mathematically, while successive operations normally cause the error to grow toward the modulus boundary, this refresh operation maps the current state to a new ciphertext with a fixed, minimal noise term: $c_{out} \approx f(m) + e_{fixed}$. By ensuring that $e_{fixed}$ remains small and independent of the input noise magnitude, the system can execute complex and non-linear machine learning operations of arbitrary depth without the risk of the noise overlapping with the message bits and causing wrong ciphertext decryptions.

\subsection{Fully Homomorphic Encryption (FHE) Preliminaries}\label{subsec:prelims}

To regroup all potential notations of the main cryptographic operations of FHE, we simulate a minimized representative scenario and all the essential operations regarding data encryption, evaluation, and decryption. 
In this scenario, a client performs a large language model inference hosted on a non-trusted server while keeping the input data confidential. 
Let $\mathcal{M}$ denote the plaintext space (e.g., integer embeddings) and $\mathcal{C}$ the ciphertext space. 
The client generates a public, private, and evaluation keys $(\mathsf{pk}, \mathsf{sk}, \mathsf{ek})$ and encrypts the input embeddings\footnote{The embeddings are the result of feeding the tokenized then encoded raw input into the transformer's embedding layer} before sending them to the server.

\begin{itemize}
	\item \textbf{Client [sends]:} Holds sensitive input text and the private key $\mathsf{sk}$, encrypts input embeddings $m \in \mathcal{M}$ using $\mathsf{pk}$, then sends it to the server.
	\item \textbf{Server [receives, computes, sends]:} Hosts the LLM and receives encrypted embeddings and the evaluation key $\mathsf{ek}$, to perform homomorphic evaluation (where main inference operations), including linear layers, attention mechanisms, and activation functions, without accessing plaintext. These phases include bootstrapping and relinearization operations to adjust the increasing noise levels within the resulting ciphertexts.
	\item \textbf{Client [receives]:} Receives the evaluated ciphertexts from the server, then executes the decryption operation using $\mathsf{sk}$ to collect the output (model logits).     
\end{itemize}

The general assumption is that the server is honest-but-curious, in other words, it will faithfully execute the model but attempts to infer sensitive information from intermediate ciphertexts. Another assumption which represents attack scenarios is that the server is compromised and the malicious actor controlling it attempts to recover the plaintext(s). FHE ensures that all intermediate activations, gradients , weights, and attention representations remain encrypted with no intermediate decryption operations.

The key FHE operations mentioned in this scenario are briefly listed, in operational order, as follows (where $c_i \in \mathcal{C}$ denotes ciphertexts in the ciphertext space): \\ 

Let $\mathcal{M}$ denote the plaintext space (e.g., integers modulo $p$) and $\mathcal{C}$ denote the ciphertext space. 
FHE allows computations on encrypted data without accessing plaintext. We summarize these operations below.

\paragraph{Key Generation} The client generates the necessary keys:
\[
\mathsf{KeyGen}(1^\lambda) \rightarrow (\mathsf{pk}, \mathsf{ek}, \mathsf{sk})
\]
where $\lambda$ is the security parameter. For RLWE-based schemes, 
$\mathsf{sk} \in \mathbb{Z}_q[X]/(X^n+1)$ and 
$\mathsf{pk} = (a, b = -a \cdot \mathsf{sk} + e)$ with small error polynomial $e$.

\paragraph{Encryption} A plaintext $m \in \mathcal{M}$ is encrypted as:
\[
\mathsf{Enc}_{\mathsf{pk}}(m) \rightarrow c
\]
In RLWE:
\[
c = (c_0, c_1) = (b \cdot r + e_0 + \Delta m, a \cdot r + e_1)
\]
where $r, e_0, e_1$ are small random polynomials and $\Delta$ is a scaling factor.

\paragraph{Homomorphic Evaluation} Given $f: \mathcal{M}^k \to \mathcal{M}$:
\[
c_{\text{out}} \gets \mathsf{Eval}(f, c_1, \dots, c_k)
\]
Operations include:
\[
c_{\text{sum}} = c_a + c_b, \quad \eta_{\text{sum}} \approx \eta_a + \eta_b
\]
\[
c_{\text{prod}} = c_a \cdot c_b, \quad \eta_{\text{prod}} \approx \eta_a + \eta_b + \epsilon q
\]
Where $\quad \eta_{\text{sum}}$ and $\quad \eta_{\text{prod}}$ refer to the noise increase in the \texttt{Add} and \texttt{Mult} operations. The noise amplitude (how large the error term is) increases roughly proportionally to parameters that depend on the polynomial ring size (n) and the ciphertext modulus logarithm $log\ q$.

\paragraph{Relinearization} Reduces ciphertext dimension after multiplications:
\[
c_{\mathsf{relin}} \gets \mathsf{Relin}(c_{\text{prod}}, \mathsf{ek})
\]

\paragraph{Programmable Bootstrapping (PBS)}: Stands for operations introduced by \texttt{concrete\--ml} that refresh noisy ciphertexts and prevent them from exceeding the correct decryption threshold
\[
c_{\mathsf{fresh}} \gets \mathsf{PBS}(c_{\mathsf{noisy}}, g)
\]
where $g$ is a programmable function (identity or non-linear).

\paragraph{Decryption} The client recovers plaintext:
\[
m = \mathsf{Dec}_{\mathsf{sk}}(c) = \left\lfloor \frac{c_0 + c_1 \cdot \mathsf{sk}}{\Delta} \right\rceil \bmod p
\]

\paragraph{Ciphertext Packing in SIMD } Multiple ciphertexts can be packed in one global ciphertext to accelerate the homomorphic operations while being executed in a SIMD fashion:
\[
c = \mathsf{Enc}([m_1, \dots, m_\ell])
\]
enabling parallel evaluation and amortizing noise growth.

\subsection{\textsc{Llama-3}'s architecture}

\begin{figure}[ht!]
	\centering
	\includegraphics[width=0.8\linewidth]{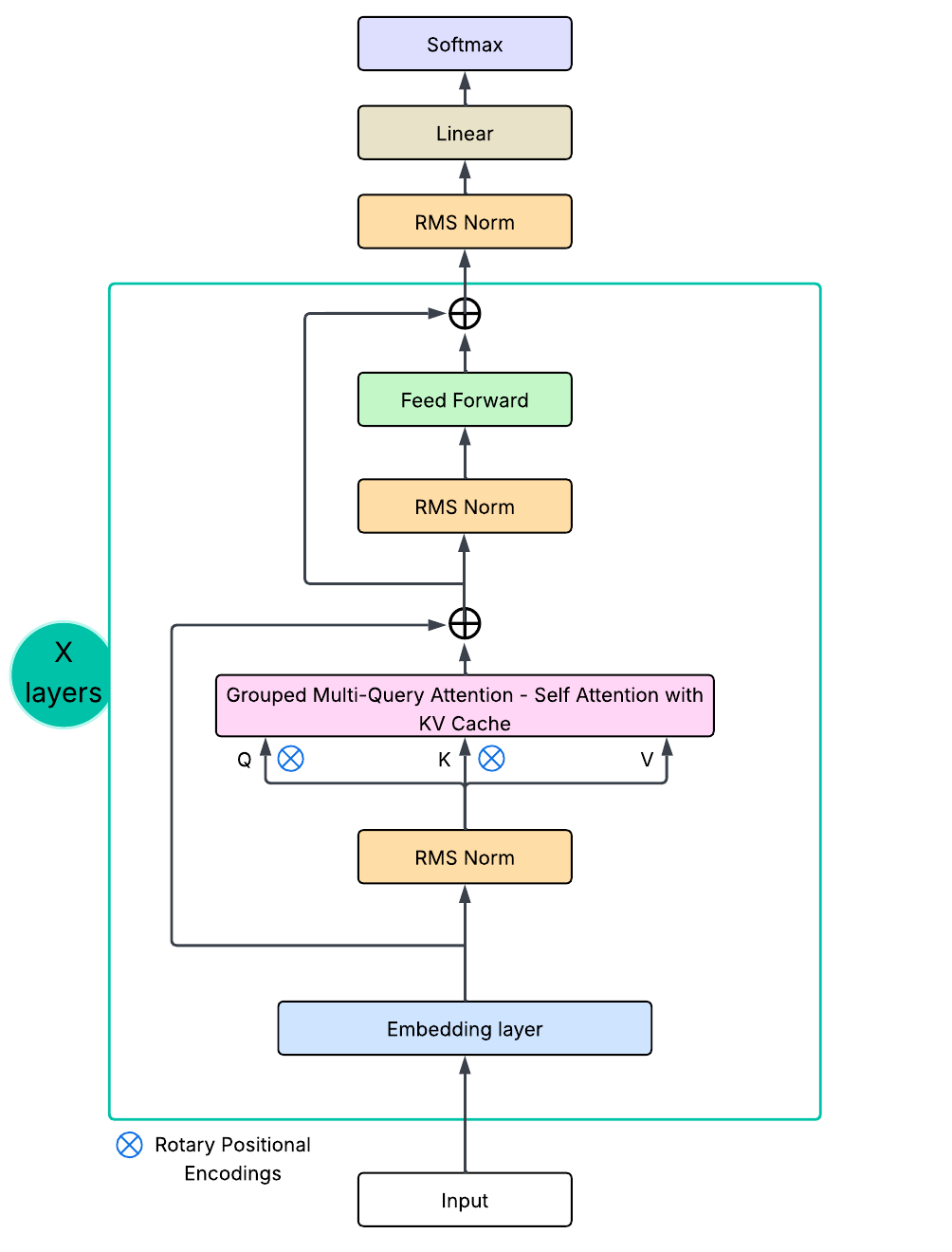}
	\caption{The \textsc{Llama-3} main architecture}
	\label{fig:llama3_arch}
\end{figure}

The \textsc{Llama-3} model's architecture is based on the transformer decoder's main design with a few changes regarding some internal functions. As Figure~\ref{fig:llama3_arch} illustrates, the embedding layer takes the raw input and transforms it into a large-dimensional vector (after tokenizing and encoding the input). Nevertheless, the main differences include the used normalization layer (or standardization, which keeps the input's values stable and within a normalized range to ensure a smooth and accelerated model training), where \textsc{Llama-3} uses the Root Mean Square normalization function (\texttt{RMS}). Additionally, the model uses the Rotary Positional Encoding method (\texttt{RoPE}) instead of the standard Positional Encoding method in the original transformer architecture. Concerning memory-related improvements in \textsc{Llama-3} for efficiency, the latter is based on Grouped Multi-Query Attention (GQA) layers to group query heads and share key-value heads within groups without affecting the model's accuracy during the inference phase. The model uses the Swish Gated Linear Units (\texttt{SwiGLU}) activation function instead of the \textit{Rectified Linear Unit} (\texttt{ReLU}) to add more non-linearity to the model, which allows adaptability towards understanding complex behaviours better than the original function. 

Some of the above-mentioned functions will be subject to transformations in \textsc{Llama-3} into a fully homomorphic encrypted inference pipeline, since they are, by nature, computationally expensive and memory demanding compared to linear functions. As HE functions introduce more computations, it is substantial to replace these functions with their efficient function twins in an FHE implementation of the \textsc{Llama-3} model.

In Section~\ref{sec:method}, we discuss leveraging an existing fully homomorphic encryption library (\texttt{concrete\--ml}) to integrate its functions within an architecturally modified \textsc{Llama-3} inference model.


\section{Methodology}\label{sec:method}

This work will inject homomorphic encryption operations (multiplications, additions, and bootstrapping) from the \texttt{concrete-ml} library into the \textsc{LLaMA-3}'s transformer architecture, requiring methods overriding and creating new classes inheriting from the mother model (will not be connected to it though under deployment setups).

We build two variants of the encrypted version \textsc{LLaMA-3}, the first one has only its first decoder layer under fully homomorphic encryption settings, while keeping the rest in plain mode. The second variant will have all its layers encrypted. The reason is to measure the impact of FHE operations on the model as whole while tracking its granular impact on a single layer. Moreover, for various applications, encrypting a subset of the model's layers might be considered a good trade-off between privacy and performance given the security and data privacy requirements and restriction bounds.

The new models will be evaluated and tested for text generation accuracy by comparing the match between the generated logits and tokens under the encrypted mode versus the usual plain model text generations.
The center of the model architecture modifications will be within its attention mechanism. Therefore, the attention function should be presented briefly before explaining the core changes in the main \textsc{Llama-3} model functions in this work.

\subsection{Attention mechanism}
The attention mechanism introduced in \cite{vaswani2017attention} has emerged as a pivotal innovation in deep learning, particularly within the domains of natural language processing, computer vision, and time-series modeling. The attention mechanism allows models to dynamically focus on relevant parts of the input sequence when producing the next word (token) while taking into consideration each element of the input. Unlike traditional sequence models such as recurrent neural networks (RNNs), which encode the entire input into a fixed-length vector, attention mechanisms provide a flexible means of accessing the entire input representation at each decoding step. There are multiple variants of the self-attention mechanism, such as Multi-Head Attention (MHA), Multi-Query Attention (MQA),  Grouped Multi-Query Attention (GQA) \cite{ainslie2023gqa}, and Sliding Window attention \cite{beltagy2020longformer}. 

At its core, the attention mechanism computes a weighted sum of input features (often referred to as values), where the similarity between a query vector and a set of key vectors determines the weights. Formally, given a query $Q$, a set of keys $K$, and corresponding values $V$, the attention output is computed as:

\begin{equation}
	\label{eq:attention_function}
	\text{Attention}(\textbf{Q, K, V}) = \text{softmax}\left(\frac{\textbf{QK}^\top}{\sqrt{(d_{emb})}}\right)\textbf{V}
\end{equation}

Where $d_{emb}$ is the dimensionality of the keys ($K$) or similarly, the embedding layer's vector dimension. This formulation, introduced in the seminal Transformer architecture \cite{vaswani2017attention}, is known as scaled dot-product attention. It enables the model to capture dependencies across different parts of the sequence regardless of their distance, thereby mitigating the limitations of fixed-context or sequential processing as viewed in RNNs.

As demonstrated in Equation~\ref{eq:attention_function}, the attention mechanism already comprises a significant amount of multiplications (dot-product \& matrix-matrix multiplications) between $Q$ and $K^\top$, then scaled by $\sqrt{d_{emb}}$ and finally multiplied with the $V$ matrix after applying the $\mathsf{softmax}$ function. Even though the amount of multiplication operations is reduced with multi-head or Grouped Multi-Query attention variants, the complexity of the attention process remains expensive. 

\subsection{FHE into \textsc{Llama-3}}

To introduce FHE in LLMs, specifically to a \textsc{Llama-3} model, we re-implement the inference decoder-based model while integrating \texttt{concrete-ml}'s main functions into the attention layers, allowing data encryption and fully homomorphic evaluation of the inference process. Therefore, the main operations and transformations within attention heads will be discussed to simplify the understanding process of our changes.

As mentioned previously, \textsc{Llama-3} uses grouped multi-query attention layers to process the inputs through a series of matrix multiplications where the input vectors are multiplied with the model weights to form the $Q$, $K$, and $V$ matrices, form the Key-Value ($\mathsf{k}$-$\mathsf{v}$) heads (which takes fewer heads number, for instance, in \textsc{Llama-3} there are $32$ heads, the ($\mathsf{k}$-$\mathsf{v}$) grouped heads will be $8$), then share them across the Query heads in the grouping process, in which each Key-Value group will be assigned to a query head projection. Next, the query matrices are multiplied by the transpose of key matrices, scaled using the square root of the embedding dimension, and then finally, multiplied by the value matrices after applying the softmax function. This operation (attention calculation) is repeated for each layer in the Llama model.

\begin{figure}[H]
	\centering
	\includegraphics[width=0.55\linewidth]{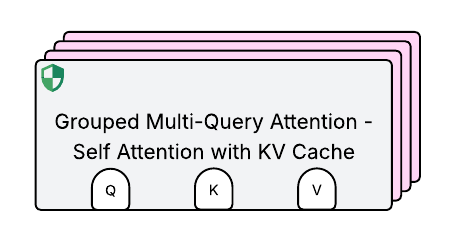}
	\caption{An abstract view of the developed \texttt{QLlamaSingleHeadAttention} implementing the main QLlamaAttention that overrides the original \textsc{Llama-3} self-Attention layers. The figure shows a single homomorphically encrypted grouped multi-query attention head}
	\label{fig:single_qllama}
\end{figure}

\begin{figure}[H]
	\centering
	\includegraphics[width=0.55\linewidth]{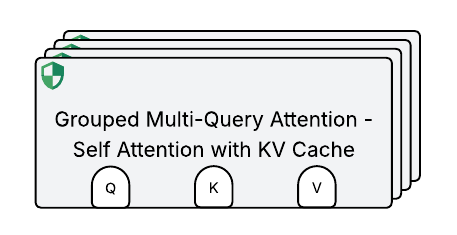}
	\caption{An abstract view of the developed \texttt{QLlama3MultiHeadsAttention} implementing the main QLlamaAttention that overrides the original \textsc{Llama-3} self-Attention layers. The figure shows a a multi-head homomorphically encrypted grouped multi-query attention heads. It is the result of concatenating multiple encrypted single attention heads.}
	\label{fig:mha_qllama}
\end{figure}

\begin{figure}[t]
	\centering
	\includegraphics[width=1.0\linewidth]{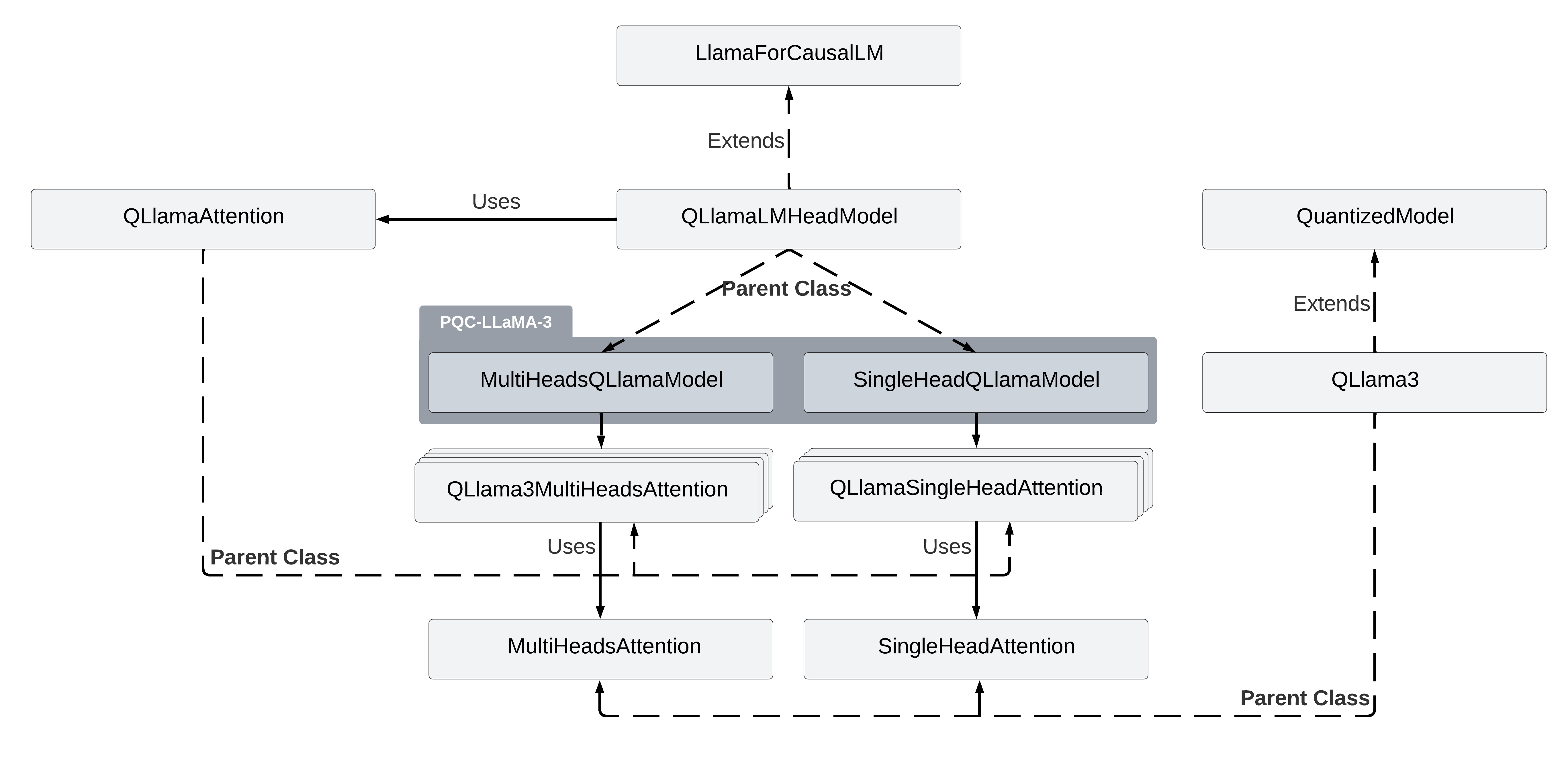}
	\caption{Class hierarchy of the proposed FHE-enabled QLLaMA-3 models. Encrypted attention modules inherit from the original LLaMA-3 architecture while selectively replacing attention heads with quantized and FHE-compatible counterparts. The classes in darker grey represents the main executable model variants of \texttt{PQC-LLAMA-3}.}
	\label{fig:class_diagram}
\end{figure}

We transform the \textsc{Llama-3} architecture into a fully homomorphically encrypted (FHE)-compatible model for inference through quantization and selective FHE-aware computation. The central aim of this approach is to adapt the attention mechanism in the \textsc{Llama-3} model such that a subset of the computations, especially the most sensitive and computationally expensive functions, can be executed securely over encrypted data using the \texttt{concrete-ml} library. To this end, we introduce two key architectural modifications (see Figure~\ref{fig:class_diagram}): the Quantized Llama Single Head Attention (\texttt{QLlamaSingleHeadAttention}) and the Quantized Llama Multi-Head Model (\texttt{QLlamaLMHeadModel}) classes, collectively enabling grouped multi-query encrypted attention heads within a \textsc{Llama-3} modified inference model.

\subsection{Quantized Attention with FHE Integration}

At the core of the proposed methodology is the re-engineering of the Llama attention layers. Standard Llama uses a multi-head attention mechanism with high-precision floating-point matrix operations. The \texttt{concrete-ml} library does not fully support multiplication and addition of floating-point numbers. Therefore, we leveraged the model integer quantization within the library on the modified \textsc{LLaMA-3} model and ran it in FHE-compatible functions. Consequently, we name our new fhe-supported and Quantized \textsc{Llama-3} model the \texttt{PQC-LLaMA-3} model. The quantized attention replaces the original attention heads with fixed-point arithmetic within this model. Specifically:
\begin{enumerate}
	\item Projection Replacement: In the SingleHeadAttention class illustrated in Figures~\ref{fig:single_qllama} and \ref{fig:class_diagram}, the initial linear transformations that generate the Query, Key, and Value vectors—originally computed via high-dimensional matmul over the attention weights (\texttt{c\_attn})—are replaced with quantized linear projections implemented using the DualArray abstraction (a proposed data structure used in \texttt{concrete-ml} library), maintaining both floating-point and integer representations.
	\item Selective Head Quantization \& Encryption: Rather than performing FHE computations across all 32 heads of \textsc{Llama-3}, we isolate a single attention head (typically the first) for encrypted evaluation (attention mechanism computations), implemented via \texttt{QLlamaSingleHeadAttention}. This balances between the desired security and model's text generation efficiency for certain applications that counts a single layer encryption sufficient to provide reasonable privacy levels for the input data and the model weights.
	\item Attention Computation Replacement: The softmax-scaled dot-product attention (i.e., $Softmax(QK^T / \sqrt{(d_{emb}})) \times V)$ is restructured in the quantized attention module. The computation is decomposed into FHE-compatible operations: encrypted dot products for $QK^T$, normalization via polynomial approximations (replacing the softmax function with a FHE-compatible approximation function), and secure matrix-vector multiplications for the attention output.
	\item FHE Execution Modes: To facilitate FHE integration at different development stages, the \texttt{QLlamaAttention} class introduces a FHE mode toggle ("disable", "simulate", or "execute"). These modes are originally offered by the \texttt{concrete-ml} developers for simulating or executing the homomorphic operations over the modified inference model (\texttt{PQC-LLaMA-3}). This allows running a quantized model on clear, non-encrypted data to quickly estimate its accuracy and performance in a Fully Homomorphic Encryption operations without the significant overhead of actual FHE computation ("simulation"). It also offers the fully homomorphic data encryption and operations via the ("execute") mode. The ("disable") mode stands for keeping the model without quantization nor FHE encryption.
\end{enumerate}

\subsection{\textsc{Llama-3} Model Modification for Quantized Integration}

The \texttt{QLlamaLMHeadModel} class extends \texttt{LlamaForCausalLM}, substituting the attention module in a designated transformer layer with our quantized version. We localize encryption overhead and maintain compatibility with downstream decoder components by targeting a specific layer (e.g., the first).

Prior to FHE compilation, the model performs a cleartext calibration pass to determine the quantization parameters required for encrypted execution. This calibration step estimates scaling factors, zero-points, and value ranges for each quantized operation, which are stored using \texttt{concrete\--ml}’s DualArray structure. These parameters ensure numerical consistency between cleartext and encrypted inference, and define the fixed-point representation that will be used within the FHE circuit.

\begin{figure}
	\centering
	\includegraphics[width=1.05\linewidth]{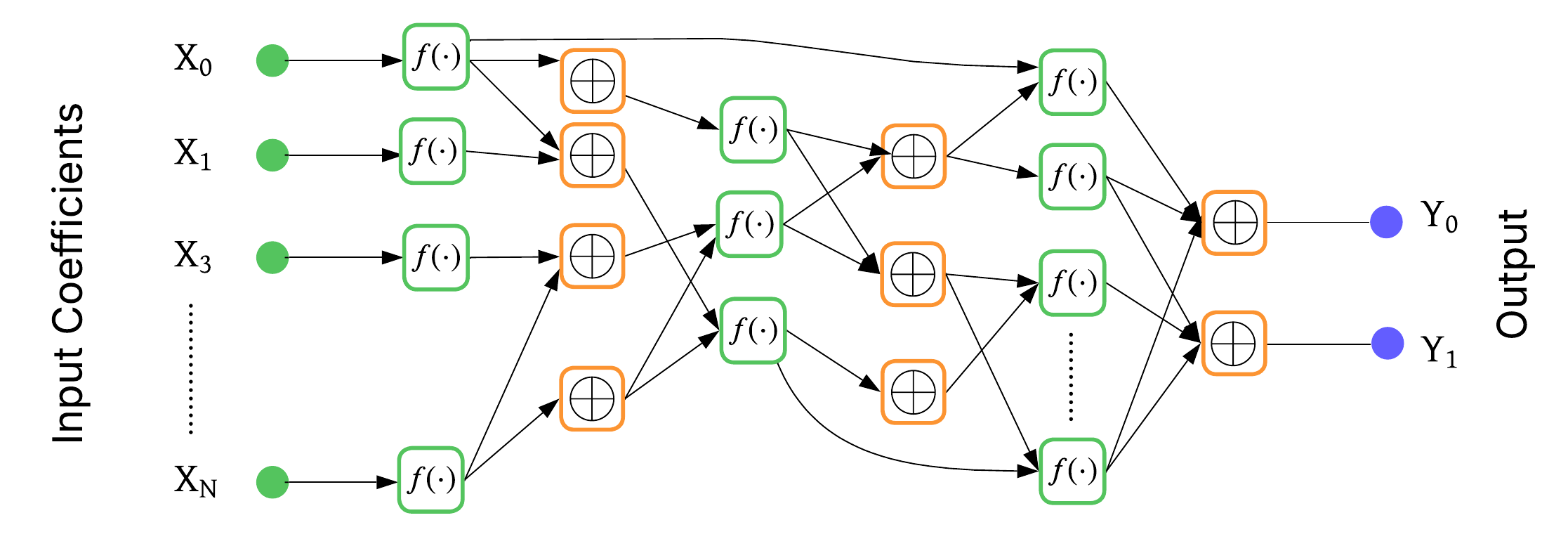}
	\caption{Programmable Bootstrapping Evaluation Circuit formed of univariate and linear functions. It reduces noise per function (Evaluation + Bootstrapping)}
	\label{fig:fhe_circuit}
\end{figure}

Following calibration, the encrypted computation graph is generated, as illustrated in Figure~\ref{fig:fhe_circuit}, by compiling the forward pass of the attention module using the \texttt{concrete\--ml} compilation API. During this phase, the FHE circuit structure is fully determined, including the placement and configuration of PBS operations. Although PBS is not executed at compile time, its count, location, and cryptographic parameters (e.g., key-switching and precision constraints) are statically inferred from the circuit graph. In particular, PBS nodes are inserted wherever nonlinear operations or noise-refresh steps are required to maintain correctness under FHE constraints. 
As a result, the compiled circuit encodes a noise-safe execution plan, while the actual noise management and bootstrapping are performed during encrypted execution, transparently to the user.

The described architecture is modular, allowing for straightforward substitution of additional layers or expansion to multi-head encrypted attention. In our implementation, in addition to \texttt{QLlamaSingleHeadAttention}, which encrypts only one head, the framework supports generalization to multiple heads through multiple duplications of the developed \texttt{SingleHeadAttention} structure, resulting in the \texttt{QLlama3MultiHeadsAttention} (See Figure~\ref{fig:mha_qllama}  compared to Figure~\ref{fig:single_qllama}).

\subsection{Homomorphic Evaluation and Llama Inference}
After the compilation step where the general FHE circuit for the attention mechanism is built, the input text (under its embedding format) is encrypted into ciphertexts using the previously generated public key (compilation phase) on the client side. Furthermore, linear layers (projections such as \texttt{q\_proj}, \texttt{k\_proj}, \texttt{v\_proj}, and \texttt{o\_proj}), quantized activations, and intermediate integer arithmetic are translated into low-level cryptographic operations, mainly additions, ciphertext multiplications, and PBS operations (for noise refresh).
Once the input is encrypted and the evaluation keys are generated, they are passed to the server side to run the expensive attention layer(s) mechanism operations in FHE mode. The server executes the circuit, producing encrypted outputs, which the client then decrypts to recover integer representations. It would result in a correct decyrption for two reasons: \textbf{\textit{i}}) the public, private, and evaluation keys were generated by the authenticated and legitimate client, \textbf{\textit{ii}}) the PBS operations ensure that the ciphertexts' noise level does not exceed the correct decryption threshold. The result decryption in integer format is then dequantized back to floating numbers.

\begin{figure}
	\centering
	\includegraphics[width=0.8\linewidth]{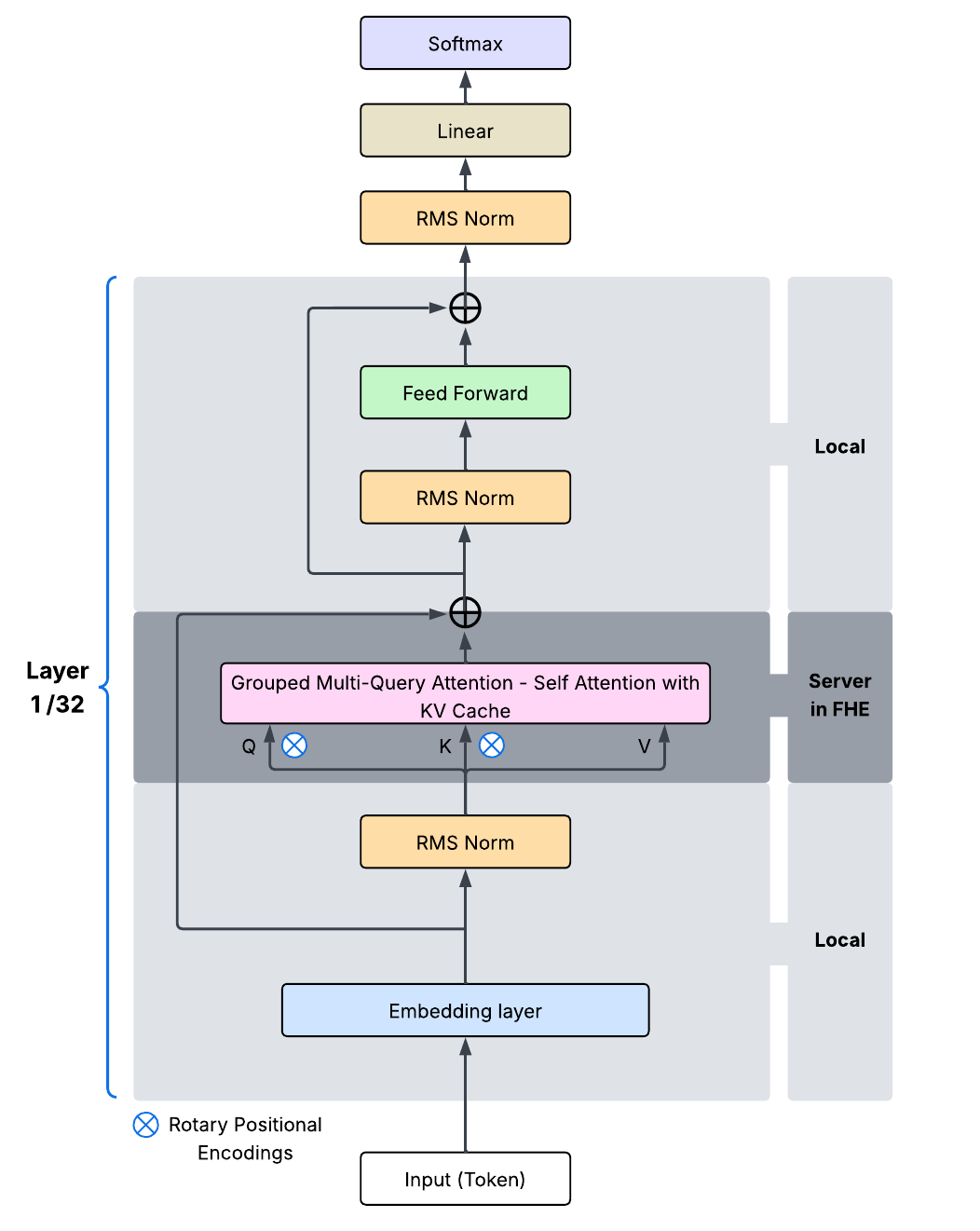}
	\caption{An overview of the \textsc{PQC-LLaMA-3} model for inference. It processes the input until the first layer locally, then sends it to the server side to homomorphically evaluate the attention mechanism. The evaluation result is sent back to the local machine to continue the inference process. This is a model describing an efficient use case of our model within a client-server model where the server has powerful hardware to run the expensive homomorphic operations on the attention mechanism.}
	\label{fig:fhe_inf}
\end{figure}

The decrypted logits (or outputs) are passed to softmax layer (not the attention softmax function, check Figure~\ref{fig:fhe_inf}) in clear, decode token IDs into words to complete the model's auto regressive text generation task~\footnote{Autoregressive generation: feed predicted token(s) back for next token generation}.

The \texttt{SingleHeadQLlamaModel} and \texttt{MutliHeadsQLlamaModel} represent the \texttt{PQC-LLaMA-3} model we propose in this work, they are the main variants in which we observe the effect of encrypting one transformer layer (and its attention heads) or all layers on the overall model's performance, including inference time and tokens/second metrics. The \texttt{SingleHeadQLlamaModel} model does not mean running it with only one layer, instead, encrypting one layer, while keep the rest ($31$ layers) in plain.

\section{Experimental Setup and Results}\label{sec:exp}
We measure the performance of our developed \textsc{PQC-LLaMA-3} model by comparing its text generation with a plain \textsc{LLaMA-3} model. Therefore, we selected a set of $79$ text prompts and feed it to each model (plain and encrypted) and for \texttt{SingleHeadQLlamaModel} and \texttt{MutliHeadsQLlamaModel} models, then incentivize it to complete the inserted prompts (unfinished sentences and paragraphs) by generating the potential next words (tokens). Furthermore, we vary the number of new tokens to generate for both encrypted model variants (\textit{single} and \textit{multi-head}) and repeat them $5$ times, then report their average.

The main metrics to evaluate in this work are the correctness of the large language model text generation in a fully homomorphic encrypted modified and quantized \textsc{Llama-3} inference model. Hence, we assess the effect on the model's accuracy after converting floating numbers to integers and run the attention mechanism arithmetic operations in FHE mode. The other metrics are the execution times of the inference and compile processes, in addition to the tokens per second readings of our proposed model compared to the plain \textsc{Llama-3} model. We used the machines described in Table~\ref{tab:hardware-specifications} for the designed trials then compared their performances. 

\begin{table}[!ht]
	\setlength{\abovecaptionskip}{2pt}
	\centering
	\caption{Hardware Specifications of both machines, running on an Ubuntu 22.04.4 LTS operating system.}
	\label{tab:hardware-specifications}
	\footnotesize
	\renewcommand{\arraystretch}{1.1}
	\setlength{\tabcolsep}{5.0pt}
	\begin{tabular}{@{}l|c|c|c|c@{}}
		\toprule
		\textbf{Node} & \textbf{Count} & \textbf{CPU} & \textbf{Storage} & \textbf{Cache} \\ 
		\midrule
		machine 1 (\texttt{M1}) & 1 & \parbox{1.8cm}{\centering AMD EPYC 7413 24-Cores ($2.65$ GHz)} & 4.5TB SSD & \parbox{1.8cm}{\centering \texttt{L1d}: 1.5 MiB  \\ \texttt{L1i}: 1.5 MiB \\ \texttt{L2}: 24 MiB \\ \texttt{L3}: 256 MiB} \\ 
		\midrule
		machine 2 (\texttt{M2}) & 1 & \parbox{1.8cm}{\centering 12th Gen Intel(R) Core(TM) i7-12700K ($3.6$ GHz} & 1TB SSD & \parbox{1.8cm}{\centering \texttt{L1}: 1.0 MiB  \\ \texttt{L2}: 12.0 MiB \\ \texttt{L3}: 25.0 MiB} \\
		\midrule
		machine 3 (\texttt{M3}) & 1 & \parbox{1.8cm}{\centering 14th Gen Intel(R) Core(TM) i9-14900K ($3.2$ GHz)} & 2TB & \parbox{1.8cm}{\centering \texttt{L1d}: 896 KiB  \\ \texttt{L1i}: 1.3 MiB \\ \texttt{L2}: 32 MiB \\ \texttt{L3}: 36 MiB} \\
		\bottomrule
	\end{tabular}
\end{table}

The experimental evaluation is organized into two main categories related to the length of the generated new tokens (context length). The first category considers evaluating the \textsc{PQC-LLaMA-3} model when generating from $1$ to $10$ new tokens, which is considered short-context generation and simulates applications such as classification/labelling or short question answering. The second category illustrates the model's generation for a longer context (from $70$ to $500$ new tokens) fitting applications such as report generation, classification with explanation, and text summarization. 

We measure the impact of the quantization (quantization set to $2$ bits since it was observed to achieve the best performance) and fully homomorphic encryption operations on the generation accuracy of \textsc{PQC-LLaMA-3} and its average inference time, then draw a few conclusions about the model's performance for short and long-text generation. Each category covers the three trial modes offered by \texttt{concrete-ml}, the \textit{disable}, \textit{simulate}, and \textit{execute} modes.
The obtained results are further discussed in the appendices section, where they are linked to the cache memory access patterns and the different processors' capacities.

\subsection{Short-context generation}
Short-context token generation, limited to $1\!-\!10$ new tokens, stresses the model’s ability to produce \emph{immediate outputs} when there is little temporal averaging.  
For a single step $t$, the logits are computed as
\begin{equation}
	\mathbf{z}_t = W_o \mathbf{h}_t,
\end{equation}
where $\mathbf{h}_t$ is the Transformer hidden state and $W_o$ is the output projection.  
Under FHE, each inner product in $\mathbf{z}_t$ is evaluated on ciphertexts $c_i$ with inherent noise $\eta_i$.  
The effective noise level per token is therefore
\begin{equation}
	\mathrm{\eta}_{t_j} \approx
	\frac{\|W_o \mathbf{h}_t\|}{\|\eta_t\|}.
\end{equation}

Where $t_{j}$ represents token $j$ from the full set of generated tokens. 
Because only a few tokens are generated, the model cannot benefit from the statistical smoothing of attention over many steps; even minor quantization or decryption noise directly perturbs $\mathrm{softmax}(\mathbf{z}_t)$\footnote{The softmax function converts the generated logits from the previous steps at a specific layer into a probability distribution, indicating the model's confidence that the input belongs to a specific class or that a particular word is the most likely next word}, making this regime a stringent test of per-step numerical fidelity.

\begin{table}[htbp]
	\centering
	\scriptsize
	\caption{Generation Accuracy (\%) / Inference Time (s) for n\_bits=2 integer quantization on Three Machines for the \texttt{SingleHeadQLlamaModel} running in two modes (\texttt{simulate/execute}).}
	
		\begin{tabular*}{\textwidth}{@{\extracolsep{\fill}}c|c|c|c|c|c|c@{}}
			\hline
			\textbf{Top-k} & \textbf{Trial} & \textbf{Accuracy} & \textbf{(\texttt{M1})} & \textbf{(\texttt{M2})} & \textbf{(\texttt{M3})} & \textbf{Compile Time (s) (\texttt{M1}/\texttt{M2}/\texttt{M3})}\\
			\hline
			\multirow{2}{*}{1}  
			& Simulate & \multirow{2}{*}{47.2} & 0.472 &  0.264 &  0.239 &  8.02 / 4.420 / 3.497\\
			& Execute  &  & 0.572 & 0.266 &  0.237 & 8.02 / 4.455 / \textbf{3.471} \\
			\hline
			\multirow{2}{*}{2}  
			& Simulate & \multirow{2}{*}{61.0} & 0.490 &  0.261 &  0.241 & 8.01 / 4.412 / 3.574 \\
			& Execute  &  & 0.474 &  0.261 &  0.241 & 7.60 / 4.342 / 3.565 \\
			\hline
			\multirow{2}{*}{3}  
			& Simulate & \multirow{2}{*}{67.4} & 0.526 & 0.262 &  0.241 & 7.9 / 4.427 / 3.627 \\
			& Execute  &  & 0.588 &  0.262 & 0.24 & 8.02 / 4.328 / 3.572\\
			\hline
			\multirow{2}{*}{4}  
			& Simulate & \multirow{2}{*}{70.6} & 0.492 &  0.262 &  0.241 & 7.58 / 4.422 / 3.640 \\
			& Execute  &  & 0.523 &  0.261 & 0.236  & 7.9 / 4.377 / 3.625\\
			\hline
			\multirow{2}{*}{5}  
			& Simulate & \multirow{2}{*}{73.1} & 0.521 &  0.261 &  0.242  & 7.82 / 4.437 / 3.638 \\
			& Execute  &  & 0.525 &  0.261 &  0.24 & 8.06 / 4.353 / 3.604\\
			\hline
			\multirow{2}{*}{6}  
			& Simulate & \multirow{2}{*}{75.9} & 0.541 &  0.261 &  0.238 & 7.72 / 4.402 / 3.571 \\
			& Execute  &  & 0.615 &  0.262 &  0.24 & 8.10 / 4.300 / 3.57\\
			\hline
			\multirow{2}{*}{7}  
			& Simulate & \multirow{2}{*}{77.6} & 0.527 &  0.261 &  0.238 & 7.61 / 4.429 / 3.651 \\
			& Execute  &  & 0.549 &  0.263 & 0.239 & 8.06 / 4.332 / 3.653\\           
			\hline
			\multirow{2}{*}{8}  
			& Simulate & \multirow{2}{*}{79.2} & 0.542 &  0.263 &  0.242 & 7.63 / 4.494 / 3.633 \\
			& Execute  &  & 0.540 &  0.262 &  0.238 & 7.58 / 4.377 / 3.616\\
			\hline
			\multirow{2}{*}{9}  
			& Simulate & \multirow{2}{*}{80.9} & 0.529 &  0.262 &  0.239 & 7.82 / 4.449 / 3.614 \\
			& Execute  &  & 0.594 &  0.262 &0.237 & 7.94 / 4.352 / 3.62\\
			\hline
			\multirow{2}{*}{10} 
			& Simulate & \multirow{2}{*}{\textbf{81.7}} & 0.558 &  0.261 &  0.238 & 7.95 / 4.401 / 3.616 \\
			& Execute  &  & 0.568 &  0.261 & 0.239 & \textbf{8.555} / 4.350 / 3.618\\
			\hline
		\end{tabular*}
	
	\label{tab:llm_quant_comparison}
\end{table}

\begin{table}[htbp]
	\centering
	\scriptsize
	\caption{Generation Accuracy (\%) / Inference Time (s) for n\_bits=2 integer quantization on Three Machines for the \texttt{MultiHeadsQLlamaModel} running in two modes (\texttt{simulate/execute}).}
	
		\begin{tabular*}{\textwidth}{@{\extracolsep{\fill}}c|c|c|c|c|c|c@{}}
			\hline
			\textbf{Top-k} & \textbf{Trial} & \textbf{Accuracy} & \textbf{(\texttt{M1})} & \textbf{(\texttt{M2})} & \textbf{(\texttt{M3})} & \textbf{Compile Time (s) (\texttt{M1}/\texttt{M2}/\texttt{M3})} \\
			\hline
			\multirow{2}{*}{1}  
			& Simulate & \multirow{2}{*}{51.6} & 1.575 &  0.909 &  0.835 & 26.190 / 14.882 / 12.286\\
			& Execute  &  & 1.558 &  0.902 &  1.055 & 26.369 / 14.667 / \textbf{12.254} \\
			\hline
			\multirow{2}{*}{2}  
			& Simulate & \multirow{2}{*}{66.4} & 1.625 &  0.893 &  0.948 & 26.138 / 14.868 / 12.737\\
			& Execute  &  & 1.618 &  0.886 &  1.142  & 26.295 / 14.592 / 12.391\\
			\hline
			\multirow{2}{*}{3}  
			& Simulate & \multirow{2}{*}{72.9} & 1.542 &  0.929 &  0.942 & 26.083 / 15.040 / 12.780\\
			& Execute  &  & 1.640 &  0.899 &  1.239 & 26.295 / 14.673 / 12.785 \\
			\hline
			\multirow{2}{*}{4}  
			& Simulate & \multirow{2}{*}{76.9} & 1.601 &  0.941 &  0.954 & 26.095 / 15.148 / 12.685\\
			& Execute  &  & 1.532 & 0.883 &  1.241 & 26.178 / 14.549 / 12.832 \\
			\hline
			\multirow{2}{*}{5}  
			& Simulate & \multirow{2}{*}{80.0} & 1.652 & 0.985 &  0.958 & 26.238 / 15.219 / 12.278\\
			& Execute  &  & 1.641 &  1.08 &  1.219 & 26.264 / 15.957 / 13.044 \\
			\hline
			\multirow{2}{*}{6}  
			& Simulate & \multirow{2}{*}{82.2} & 1.653 & 0.967 &  0.958 & 26.323 / 15.330 / 12.502\\
			& Execute  &  & 1.616 &  1.0 & 1.220 & 26.213 / 15.589 / 12.958 \\
			\hline
			\multirow{2}{*}{7}  
			& Simulate & \multirow{2}{*}{83.6} & 1.554 &  0.965 & 0.953 & 26.083 / 15.425 / 12.496\\
			& Execute  &  & 1.631 &  1.0 &  1.232 & 26.108 / 15.656 / 12.834 \\           
			\hline
			\multirow{2}{*}{8}  
			& Simulate & \multirow{2}{*}{85.8} & 1.600 &  0.939 & 0.947 & 26.284 / 15.252 / 12.794 \\
			& Execute  & & 1.616 &  0.970 & 1.259 &  26.170 / 15.608 / 12.733 \\
			\hline
			\multirow{2}{*}{9}  
			& Simulate & \multirow{2}{*}{86.8} & 1.629 &  0.957 & 0.956 & \textbf{26.343} / 15.407 / 12.809 \\
			& Execute  &  & 1.599 &  0.905 & 1.244 & 26.329 / 15.056 / 12.773 \\
			\hline
			\multirow{2}{*}{10}  
			& Simulate & \multirow{2}{*}{\textbf{87.4}} & 1.679 &  0.932 & 0.956 & 26.088 / 15.321 / 12.554 \\
			& Execute  &  & 1.635 & 0.904 &  1.261 & 26.295 / 15.083 / 12.779\\
			\hline
		\end{tabular*}
	
	\label{tab:llm_quant_comparison_MH}
\end{table}



Tables \ref{tab:llm_quant_comparison} and \ref{tab:llm_quant_comparison_MH} illustrate the multiple runs of the \texttt{SingleHeadQLlamaModel} and \texttt{MutliHeadsQLlamaModel} models for top-$k$ $\in$ $[1,10]$ tokens across all available fhe modes.

\subsubsection{\textbf{Single-Head FHE-model performance}}

The results for the \texttt{SingleHeadQLlamaModel} are summarized in Table \ref{tab:llm_quant_comparison}.
Across all machines, the generation accuracy steadily increased with higher top-$k$ values, starting from 47.2\% for $k=1$ to 81.7\% for $k=10$, following the expected top-$k$ sampling trend.
Interestingly, quantization did not degrade accuracy as simulated and executed modes preserved identical generation scores as the unquantized baseline, confirming the correctness of the 2-bit integer quantization scheme.

From a performance standpoint, inference times on \texttt{M3} were consistently the lowest across all settings, with average inference latencies, for \texttt{simulate/execute} modes, around $0.24/0.238$s compared to $0.261/0.262$s on \texttt{M2} and $0.519/0.554$s on \texttt{M1}.
This reflects the superior single-thread performance of the Intel i9-14900K compared to the other processors given their turbo frequencies, $6$ GHz for \texttt{M3}, $5$ GHz for \texttt{M2}, $3.6$ GHz for \texttt{M1}, respectively.

Compilation times showed a similar pattern, decreasing from $8.55$ s on \texttt{M1} to $3.47$ s on \texttt{M3} (see the row in bold in Table~\ref{tab:llm_quant_comparison}). Only the maximum values of the compiling times were considered.

This reduction aligns with the faster memory access and higher turbo frequencies of consumer CPUs, suggesting that CPU frequency and memory latency are stronger determinants of compilation speed than total threads or system RAM size.

Comparing the quantization modes, both “\texttt{simulate}” and “\texttt{execute}” introduced negligible runtime overhead (under 5\%) relative to the unquantized baseline.
This indicates that integer-based quantized operations were efficiently integrated within the CPU execution pipeline, confirming the low-bit quantization can be CPU-friendly with minimal computational penalty.

\subsubsection{\textbf{Multi-Head FHE-model performance}}

The \texttt{MultiHeadsQLlamaModel} results, presented in Table \ref{tab:llm_quant_comparison_MH}, follow similar accuracy trends compared to Table~\ref{tab:llm_quant_comparison}, but with overall higher performance.
Accuracy improved from 51.6\% ($k = 1$) to 87.4\% ($k = 10$), confirming that the multi-head configuration better captures contextual variance even under low-bit quantization. However, the additional attention heads increased computation time. Average inference latency (\texttt{simulate/execute}) rose to approximately $1.611/1.60$s on \texttt{M1}, $0.941/0.936$s on \texttt{M2}, and $0.94/1.211$s on \texttt{M3}.

Since all computations were performed on CPUs, this increase stems from the higher volume of matrix multiplications and attention aggregation steps required per inference cycle. Nevertheless, the latency growth, compared to the single-head quantized model, remained sublinear relative to the number of heads, showing that multi-threaded CPU execution effectively distributed the operations across cores.

Compilation times for the multi-head model were notably longer, the maximum timing reached $26.343$s on \texttt{M1}, while the minimum compilation time was took $12.254$s on \texttt{M3}, due to the additional complexity in generating optimized kernel graphs for multi-head attention paths.
Once again, compile times correlated strongly with CPU clock frequency and memory access speed, reaffirming that CPU-bound quantized models remain sensitive to low-level architectural characteristics such as cache size and instruction-level parallelism. These and subsequent results are discussed in depth in Section~\ref{sec:cache}.

\subsection{Long-context generation}

In long-context generation the model autoregressively (where the generation of the next token depends also on the previously generated ones) produces up to $500$ tokens, repeatedly updating each model's decoder layer hidden states.
This setting evaluates the \emph{cumulative stability} of the encrypted pipeline, the ciphertext noise effect along with homomorphic operations controlling error across layers.  
We evaluate the models within the long-context generation for both \texttt{PQC-LLaMA-3} model's variants, the \texttt{SingleHeadQLlamaModel} and \texttt{MultiHeadsQLlamaModel} models.

\vspace{2.0mm}
\subsubsection{\textbf{Single-Head FHE model performance}}

For the \texttt{SingleHeadQLlamaModel}, accuracy improved monotonically from 91.3\% at top-$k$ = 70 to 98.2\% at top-$k$ = $500$ (see Table~\ref{tab:SHLong}).
The increase reflects the model’s enhanced capacity to consider broader token distributions during sampling.

The results reveal a clear inverse relationship between CPU frequency and latency, with \texttt{M3} achieving the lowest times. The inference time differences between \texttt{M2} and \texttt{M3} ($\sim10$\%), even though CPU frequency-wise, \texttt{M2} should be performing better than \texttt{M3}, can be attributed to the L2/L3 cache size and memory-controller efficiency, rather than to thread count.
\texttt{M1}’s higher latency mainly stems from its lower per-core frequency, which introduces additional data-access delays in FHE encryption/decryption cycles.
The stability of inference times (standard deviation $<5\%$) also suggests that FHE-enabled quantized operators were efficiently optimized for CPU cache hierarchies, achieving near-deterministic performance.
Overall, the \texttt{SingleHeadQLlamaModel} demonstrates that FHE-encrypted inference can achieve sub-second latency on modern CPUs.

\vspace{2.0mm}
\subsubsection{\textbf{Multi-Head FHE model performance}}

As Table~\ref{tab:SHLong} shows, the \texttt{MultiHeadsQLlamaModel} displayed similar accuracy trends, increasing from $94.7$\% at top-$k$=$70$ to $\sim97.7$\% at top-$k$=$500$. This improvement confirms that the multi-head structure enhances contextual representation, even under a series of expensive FHE operational mode. 

As expected, the introduction of multiple attention heads increased computational cost. On \texttt{M1}, inference time varied between $1.55$ s and $1.72$ s, while \texttt{M2} recorded values between $1.15$ s and $2.07$ s, and \texttt{M3} achieved the lowest latency between $0.79$ s and $1.02$ s. The results again demonstrate that inference time is primarily determined by per-core clock frequency and cache hierarchy efficiency. The higher variance observed on \texttt{M2} at larger top-$k$ values indicates occasional cache saturation or operating system context-switching overheads, as multiple encrypted tensors compete for limited \texttt{L2} and sometimes a generous \texttt{L3} cache space.

Compared to the single-head variant, the multi-head model incurred an average latency increase of approximately $3.5 \times$ on \texttt{M1}, $4.3 \times$ on \texttt{M2}, and $4.3 \times$ on \texttt{M3}. This moderate increase demonstrates that quantized multi-head FHE inference remains computationally feasible on high-frequency CPUs, especially when parallelized across available cores. The latency scaling was sublinear relative to the number of attention heads, confirming that the parallelization of encrypted operations effectively mitigates the expected computational overhead.

\begin{table}[htbp]
	\centering
	\caption{Generation Accuracy (\%) / Inference Time (s) for n\_bits=2 quantization on Three Machines for \texttt{SingleHeadQLlamaModel} / \texttt{MultiHeadsQLlamaModel} models running in the FHE mode (\texttt{execute}).}
	
			\begin{tabular*}{\textwidth}{@{\extracolsep{\fill}}lccccc@{}}
				\hline
				\textbf{Top-k} & \textbf{Accuracy} & \texttt{M1} & \texttt{M2} & \texttt{M3} \\
				\hline
				70  & 91.3 / 94.7 &  0.556 / 1.553 &  0.256 / 1.478 & 0.233 / 0.790 \\   
				80  & 91.7 / 94.8 &  0.501 / 1.545 & 0.258 / 1.403 & 0.237 / 0.879 \\  
				90  & 92.1 / 94.9 &  0.508 / 1.547 & 0.256 / 1.433 & 0.235 / 0.969 \\
				100 & 92.3 / 95.1 &  0.482 / 1.591 & 0.257 / 1.260 & 0.235 / 0.938 \\
				110 & 92.4 / 95.1 &  0.478 / 1.643 & 0.257 / 1.154 & 0.237 / 0.841 \\				
				120 & 92.8 / 95.3 &  0.464 / 1.646 & 0.261 / 1.248 & 0.234 / 0.926 \\
				300 & 96.2 / 97.1 & 0.456 / 1.633 & 0.268 / 1.762 & 0.235 / 1.02 \\
				400 & 96.5 / 97.4 & 0.487 / 1.721 & 0.270 / 1.899& 0.236 / 0.928\\
				500 & \textbf{98.2} / \textbf{97.7} & 0.516 / 1.652 & 0.276 / 2.070 & \textbf{0.236} / \textbf{0.825} \\
				\hline
			\end{tabular*}
		
		\label{tab:SHLong}
\end{table}

\subsection{Analysis of Long- versus Short-Context Encrypted Generation}

The objective of these experiments is to examine how fully homomorphic encryption and quantization affect text generation performance in the \textsc{PQC-LLaMA-3} model under two distinct regimes: (i) \emph{short-context} generation of 1–10 new tokens, and (ii) \emph{long-context} generation, extending up to 500 new tokens.  
Although both settings employ identical model weights, encryption parameters, and keys, they reveal different operational sensitivities.  
The Short-context emphasizes the immediate impact of encryption and quantization, including ciphertext noise on each prediction step, while long-context generation probes the cumulative noise behaviour and the stability of encrypted arithmetic over extended autoregressive sequences. 

Two main factors affect the \texttt{PQC-LLaMA-3} variants (\texttt{SingleHeadQLlamaModel} and \texttt{MultiHeadsQLlamaModel}) performance are the noise introduced by encrypted data and operations, in addition to the ciphertext packing mechanism in \texttt{concrete\--ml} library. Their contribution to the performance (positively or negatively) is discussed briefly in the next two subsections. 

\subsubsection{Noise under Encrypted Operations}
The base architecture of \textsc{LLaMA-3} is a decoder-only Transformer with $L$ layers ($L=32$), each containing GQA self-attention, feed-forward sublayers, layer normalization, and residual connections.  
At decoding step $t$, the hidden representation $\mathbf{h}_t$ is updated as in the standard Transformer formulation, and token probabilities are obtained via
\begin{equation}
	P(x_t \mid x_{<t}) = \mathrm{softmax}(W_o \mathbf{h}_t),
\end{equation}
where $W_o$ denotes the output projection matrix, and the softmax represents the decoder layer's final function.  
In the \textsc{PQC-LLaMA-3} model, these floating-point operations are replaced by fixed-precision arithmetic over ciphertexts encrypted in the RLWE domain, introducing \emph{quantization} and \emph{homomorphic noise}.

Quantization error arises from mapping real/floating point-valued tensors to low-bit-width integer representations prior to encryption (using 2-bit quantization since \texttt{concrete\--ml} support only integer arithmetic operations). Model parameters and intermediate activations are scaled and rounded according to calibrated quantization parameters, ensuring compatibility with homomorphic evaluation while bounding numerical distortion. This error is deterministic and fixed once the quantization scheme is selected.

Homomorphic noise, by contrast, is an inherent property of RLWE-based fully homomorphic encryption schemes. Each ciphertext embeds a noise component that guarantees semantic security (see Sub-section~\ref{subsec:math}). During encrypted inference, this noise grows with every homomorphic operation—most notably ciphertext–ciphertext multiplications used in dot-product computations within the attention mechanism. Additions incur negligible noise growth, while multiplications result in multiplicative noise expansion proportional to the ciphertext modulus and operand magnitudes.

These noise factors decreases the encrypted model's accuracy especially when the noise levels exceeds the calculated budget, leading to wrong decryptions of homomorphically calculated attention values, which would consequently, lead to wrong next token generation. 

To prevent uncontrolled noise accumulation and ciphertext invalidation, the \textsc{PQC-LLaMA-3} compilation process explicitly inserts PBS operations at strategically selected points in the encrypted attention circuit. PBS simultaneously refreshes ciphertexts by reducing accumulated noise and enables the evaluation of non-linear functions (e.g., activation and non-linear functions approximations) in the encrypted domain. Importantly, the placement and frequency of PBS operations are determined statically during circuit compilation using the library's framework, based on worst-case noise growth analysis of the computation graph.

\subsubsection{Ciphertext Packing}
In the FHE setup, ciphertext packing embeds multiple ciphertext slots into a single ciphertext via the Chinese Remainder Theorem (CRT) for batching, and Number Theoretic Transforms (NTT) for efficient polynomial arithmetic, enabling SIMD-style parallel and fast matrix multiplications within the attention and feed-forward layers.  
This design amortizes noise across slots and reduces per-slot variance, effectively improving the arithmetic precision of batched operations. 
Following ciphertext–ciphertext multiplications, key switching (or relinearization) re-expresses results under the same secret key and controls modulus expansion, mitigating excessive noise accumulation (which is achieved by the PBS operations).

These mechanisms ensure the numerical stability of multi-layer encrypted computation, particularly over longer generation spans, which interprets the slightly increasing accuracy for the chosen top-$k$ values (see Table~\ref{tab:SHLong}) even with the presence of FHE operations and the augmenting noise effects.

Empirically, short-context generation exhibits a sharper sensitivity to quantization precision and single-step ciphertext noise, as predictions depend heavily on the most recent encrypted hidden state.  
This results in a lower Effective Precision Ratio (EPR) at each decoding step:
\begin{equation}
	\mathrm{EPR}_{\text{short}} \sim \frac{\|W_o \mathbf{h}_t\|}{\|\eta_{\text{short}}\|},
\end{equation}
where $\eta_{\text{short}}$ denotes the noise component in the encrypted logits.  

In contrast, during long-context generation, ciphertext packing and periodic key switching distribute and stabilize noise across multiple tokens, leading to a higher effective EPR:
\begin{equation}
	\mathrm{EPR}_{\text{long}} \sim \frac{\|W_o \mathbf{H}\|}{\|\eta_{\text{long}}\|_\text{eff}},
\end{equation}
where $\mathbf{H}$ collects the hidden states of all generated tokens.  
Experimentally, this behaviour manifests as improved generation coherence and reduced degradation in perplexity for sequences up to 500 tokens.  
The results suggest that noise averaging through ciphertext packing and recurrent key switching contributes to better overall numerical stability, explaining why the \textsc{PQC-LLaMA-3} model maintains higher accuracy during long-context decoding despite fully homomorphic constraints.

\subsection{The compilation bottleneck}
As observed in Tables~\ref{tab:llm_quant_comparison} and \ref{tab:llm_quant_comparison_MH}, the model compilation process to generate the FHE circuit forms a bottleneck for this work since the operation execution times exceeds $8$ secs and $26$ secs for the \texttt{SingeHeadQLlamaModel} and \texttt{MultiHeadsQLlamaModel} models, respectively. While it appears to be an unsatisfying result where a user would not possibly wait for multiple seconds to have a text generated, nevertheless, this is avoidable by simply constructing the model's FHE circuit only once, then apply it on the ensemble of prompts/queries the users intend to submit. The circuit can be changed or re-constructed in case the task's nature changes or when the prompts' structure, type, and length vary due to multiple querying scenarios. 

The potential re-generation of the FHE circuit is enabled by the flexibility of the developed \textsc{PQC-LLaMA-3} model in this work, it offers model compilation as callable function at any desired code segment or model inference routine.

\subsection{The KV-Cache effect}
As noticed in Tables \ref{tab:llm_quant_comparison}, \ref{tab:llm_quant_comparison_MH}, and \ref{tab:SHLong}, the average inference times are almost the same regardless of the top-$k$ value (increasing number of generated new tokens). It seems logical to expect rising inference times whenever the model generates more text, however, this performance is valid and achievable thanks to the \textit{KV-caching} technique implemented in \textsc{LLaMA-3} models, and consequently, inherited in our \textsc{PQC-LLaMA-3} model as well. This option enables the auto regressive model to avoid re-computing the previous sequence of tokens preceding the one to be generated. It is an expensive computational operation to perform the same operations repeatedly, therefore, \textit{KV-Caching} technique allows the model to store the previously generated sequence in the cache and fetch it for future generations. 

To delve deeper into investigating this specific execution and memory behaviour of the model, we collected microarchitectural performance metrics using the Linux \texttt{perf} tool to quantify the execution behaviour of each model variant (\texttt{single} and \texttt{multi} heads models) across different top-$k$ configurations. To ensure consistency and reduce noise, each experiment was executed on a CPU core pinned to a single logical processor using \texttt{taskset}, avoiding thread migration. 

Hardware performance counters were recorded using \texttt{perf stat} with a fixed set of events (their names may vary depending on the processor and its kernel architecture), including \texttt{cycles}, \texttt{instructions}, \texttt{L1-dcache-loads}, \texttt{L1-dcache-load-misses}, \texttt{L1-icache-loads}, \texttt{L1-icache-load-misses}, \texttt{L2\_RQSTS.REFERENCES}, \texttt{L2\_RQSTS.MISS}, \texttt{LLC-loads}, \texttt{LLC-load-misses}. Multiple runs ($10$ trials per configuration) were averaged to account for system variability, and the collected counters were subsequently used to observe the cache memory accesses on all levels which explains the model's behaviour (similar inference times for all top-$k$ values). 
Appendix \ref{sec:cache} illustrates the results from the collected cache memory accesses, for both \texttt{SingeHeadQLlamaModel} and \texttt{MultiHeadsQLlamaModel} models, and then discusses further the enhanced memory access nature in \textsc{LLaMA-3} models which also manifests in the \textsc{PQC-LLaMA-3} encrypted model.

\subsection{FHE-related metrics}

We propose a set of new metrics to evaluate the efficiency of the \texttt{PQC-LLaMA-3} model including the number of executed PBS operations, the average of PBS count per generated token, the average needed memory per token (in bytes), and the overall throughput for both \texttt{PQC-LLaMA-3}'s variants, \texttt{SingleHeadQLlamaModel} and \texttt{MultiHeadsQLlamaModel}.

Under this FHE-secured \texttt{LLaMA-3} implementation, a sequence length of $6$ tokens (for text completion or generation) requires $5208$ PBS operations where the implementation reported in \cite{ConcreteML} requires $11622$ for a FHE \texttt{GPT-2} model. Despite that both models have different attention head architectures, it still counts as a relatively more efficient implementation compared to the mentioned prior work.

We introduce some new metrics recorded from the model's evaluation trials, illustrated in Table \ref{tab:fhe_metrics_comparison}. In order to capture the following metrics, we run trials varying the $top-k$ values within the range [$10-500$] and for each value, we run the models on the $79$ text prompts then average the recorded metrics, except for the $PBS\_count$ which is fixed for each model.

\subsubsection{Average Throughput}

The average throughput quantifies the effective inference speed of the model under encryption. It is defined as
\begin{equation}
	\mathrm{AvgThroughput} = \frac{\mathrm{AvgTokensPerSecond}}{\mathrm{AvgExecutionTime}},
\end{equation}
where AvgTokensPerSecond denotes the average number of generated tokens per second, and AvgExecutionTime corresponds to the mean end-to-end execution time per inference run. This normalized formulation mitigates variability across runs and provides a stable indicator of sustained token generation performance. In the FHE setting, throughput reflects both the cost of homomorphic arithmetic (e.g., encrypted matrix multiplications and polynomial approximations) and the overhead introduced by cryptographic primitives such as programmable bootstrapping.

\subsubsection{Average PBS per Token}

The average PBS per token measures the cryptographic intensity of encrypted inference. It is computed as
\begin{equation}
	\mathrm{Avg\_PBS\_Token} = \frac{\mathrm{PBS\_Count}}{\mathrm{Generated\_Tokens}},
\end{equation}
and represents the mean number of programmable bootstrapping operations required to generate a single token. Since PBS operations dominate latency and energy consumption in RLWE-based FHE schemes, this metric provides a direct measure of the homomorphic cost per output token. Importantly, this quantity depends on the encrypted circuit structure—such as the number of attention heads, sequence length, and non-linear layers and functions—and is independent of decoding parameters such as \textit{top-k}.

\subsubsection{Average Memory per Token}

The average memory per token, expressed in bytes, captures the memory overhead induced by encrypted inference. It is defined as
\begin{equation}
	\mathrm{Avg\_Mem\_Token} = \frac{\mathrm{Total\_Mem}}{\mathrm{Generated\_Tokens}}.
\end{equation}
This metric accounts for the storage required by encrypted activations, intermediate ciphertexts, and bootstrapping artifacts throughout the inference process. Lower values indicate better memory efficiency and improved scalability, which are critical for deployment on resource-constrained hardware.

\begin{table}
	\caption{Encrypted inference efficiency metrics for single-head and multi-head PQC-LLaMA-3 models on machine \texttt{M3} given as averages across multiple top-k values for each \texttt{PQC-LLaMA-3} model. We refer to \texttt{SingleHeadQLlamaModel} as \texttt{Single} and \texttt{MultiHeadsQLlamaModel} as \texttt{Multi}} \label{tab:fhe_metrics_comparison}
	\renewcommand{\arraystretch}{1.0}
	\setlength{\tabcolsep}{2.0pt}
	\begin{tabular*}{\textwidth}{@{\extracolsep{\fill}}lcccccc@{}}
		\toprule
		& \multicolumn{3}{c}{\texttt{M1}} & \multicolumn{3}{c}{\texttt{M3}} \\ \cmidrule{2-4}\cmidrule{5-7}
		Model & Throughput & PBS/token & Mem/token & Throughput & PBS/token & Mem/token \\
		\midrule
		\texttt{Single}  & 355.30  & 60.51 & 18952.67 & 77.26  & 137.62 & 2019.96 \\
		\texttt{Multi}   & 16.67 & 9604.25 & 597.46  & 8.91 & 13067.62 & 1119.11 \\
		\bottomrule 
	\end{tabular*}
\end{table}

The results shown in Table~\ref{tab:fhe_metrics_comparison} demonstrate a decent average throughput for the \texttt{SingleHeadQLlamaModel} and an efficient distribution of PBS operations per generated token, while consuming $2.02$ KBytes of main RAM memory. As for \texttt{MultiHeadsQLlamaModel}, its average throughput is low compared to the single FHE-head model, reflected by the high average PBS operations per generated token ($13067.62$) which explains the low throughput, since a higher PBS per token would cause the model to perform poorly increasing the latency of its text generation. However, the memory consumption is nearly the half of what \texttt{SingleHeadQLlamaModel} consumes. 
It is substantial to note that the models use the cache memory and its levels besides the main memory when storing the packed ciphertexts, attention matrices, non-linear activations and gradients.

\begin{table}
	\centering
	\caption{Assessment of FHE protections for common attacks in LLM deployment.}
	
	\begin{tabular}{p{0.23\linewidth} p{0.14\linewidth} p{0.55\linewidth}}
		\toprule
		\textbf{Attack} & \textbf{FHE-protected} & \textbf{Explanation} \\
		\midrule
		Prompt / context leakage & Yes & If intermediate states and context remain encrypted and never decrypted, server-side leakage is prevented. Leakage can still occur via decrypted outputs or side channels. \\
		\addlinespace
		Output reconstruction (decryption guessing) & Yes & Properly configured FHE ciphertexts do not reveal plaintext, but implementation errors, malleability, or metadata/timing side channels can leak information. \\
		\addlinespace
		Gradient leakage (reconstruction) & Conditionally Yes & FHE protects gradients/activations \emph{in-flight} (ciphertexts). If gradients or aggregated updates are revealed in plaintext, or if protocol steps require decryption, leakage remains possible. Hybrid/training protocols may introduce leakage points. \\
		\addlinespace
		Data poisoning & Partial & Encryption does not prevent an authorized adversary from submitting poisoned data; while obscuring inputs can hinder targeted poisoning, hardening the attempts to poison data and complicates dataset auditing, however, validating inputs and queries submission authentication is required to achieve a better security. \\
		\addlinespace
		Side-channel attacks (timing, cache, EM) & No & FHE does not inherently remove physical side-channels. Constant-time implementations, hardware isolation (TEEs), and side-channel hardening remain necessary. \\
		\addlinespace
		Model extraction (stealing) & No & FHE hides internals from an entity that only observes ciphertexts, but black-box extraction via plaintext query outputs is still possible unless outputs are restricted (DP, clipping, rate limits). \\
		\addlinespace
		Adversarial example injection & No / Partial & FHE prevents internal gradient access, but black-box adversarial attacks (query-based) remain feasible. \\
		\bottomrule
	\end{tabular}
	\label{tab:qlama_attacks}
\end{table}

\section{PQC-LLaMA-3 vs. attacks on LLMs}\label{sec:qllama_llm}

We provide in Table~\ref{tab:qlama_attacks} a set of attacks on AI models and LLMs, mostly targeting decrypted outputs and weights acquisition to perform a successful attack. Nevertheless, it is substantial to set a clear assumption about the environment where our developed model would best fit. Therefore, we assume the following while providing the model's FHE protection assessments:
The cryptographic security of the system relies on the standard hardness assumptions underlying lattice-based FHE schemes (e.g., Learning With Errors). Encryption and decryption are performed by a trusted client holding the public and private keys, while the evaluation key is delivered to the server (where evaluation takes place). The FHE server operates deterministically without malicious code injection or output tampering. Model weights and quantized tensors are pre-encrypted or protected in secure memory before deployment. Output decryption and interpretation occur in a trusted environment isolated from the adversary.

Even though fully homomorphic encryption is quantum-resistant, it is not immune to all types of attacks targeting machine learning, deep learning, or large language models. There are always risks and potential security flaws surrounding the data and model's privacy, which are a direct result of bad software implementations, hardware implementations, or both. Moreover, there exists a set of attacks that target even FHE-encrypted models where the encryption cannot prevent their risks and consequences on data and model privacy (Table~\ref{tab:qlama_attacks} discusses a few of them). 

Given this assessment, we highly recommend the coupling of this proposed secure version \textsc{PQC-LLaMA-3} model with a set of other security mechanisms to reinforce the overall model's security and data privacy against the mentioned attacks. The security mechanisms comprise input sanitization and output validation, differential privacy (DP), output clipping, query limits, and establishing  trusted execution environments (TEEs).

\section{Conclusion}\label{sec:conc}
In this work of inner operation adaptation of the \textsc{LlaMA-3} model, introducing FHE operations for secure inference, we switched the main attention mechanism (Grouped Multi Query Attention) layers' operations with the FHE primitives to form a post-quantum secure inference \textsc{PQC-LlaMA-3} model where attackers would be unable to reach, distort, or steal private or sensitive information (the input prompts, the weights, and gradients), due to the hardness level of the used homomorphic encryption scheme. We run two main evaluation scenarios, short and long context text generation, while assessing the FHE model's generation accuracy, inference's low latency on the machine's CPUs, and the tokens throughput. The \texttt{MultiHeadsQLlamaModel} reached an $87.4\%$ generation accuracy in $0.904$s on a $3.6$ GHz machine (\texttt{M2}) when generating new $10$ tokens (Short-context generation), While the \texttt{SingleHeadQLlamaModel} achieves $98.28\%$ accuracy in $0.236$s  on \texttt{M3} when generating $500$ new tokens (Long-context generation). The \texttt{MultiHeadsQLlamaModel} also attains $97.7\%$ accuracy under one second ($0.825$$s$). 

The evaluation results showed that the proposed \textsc{PQC-LLaMA-3} model is capable of reaching relatively high text generation accuracies while running in fully homomorphic encryption mode and keeping a reasonable tokens per second rate ($80$ tokens/sec). Moreover, long-context generation outperforms short-context generation due to the distribution of the noise resulting from homomorphic operations on a longer input sequence, creating a diminishing noise effect on the overall accuracy. As for the FHE cost on the \texttt{PQC-LLaMA-3} model, it demonstrates a relatively efficient cost regarding the number of PBS operations and the memory required per token generation, while benefitting from the different cache memory levels achieving decent token generation throughput and latencies.

This performance proves utility for applications such as explainable AI (XAI), generating reports, summarization, and automated chatbot-based assistance, since all these applications require longer-context new context generation. Furthermore, the developed model can be used for classification tasks, illustrated by the satisfying results acquired within the short-context generation scope.

The investigated models' behaviour reaching similar inference times under multiple top-$k$ values led to the compelling cache memory fetching routines enabled by the \textit{KV-caching} technique inherited from \textsc{LLaMA-3} model. Further cache memory investigations were conducted along with the inference times, and measuring the tokens per second rates experiments results were listed in the Appendices \ref{sec:inf_times} and \ref{sec:cache}.

This work is essential and a cornerstone for advanced data privacy preservation on LLMs, it secures the model (inference pipeline) and its data while running inferences, preventing a lethal set of cyber attacks from reaching plaintext user inputs, weights, and gradients. Attacks such as prompt and context leakage, output reconstruction, gradient leakage, and data poisoning (it hinders targeted data poisoning, but to a certain level). An ultimate use case of this research work is deploying \textsc{PQC-LLaMA-3} models for inference (text generation, classification, explainability tasks) in a chain of pre-configured security and data privacy preserving settings (to provide a comprehensive secure deployment environment), focusing on preventing attackers from reaching sensitive information under any format.


\begin{appendices}
	
	\section{Inference times \& Tokens per Second}\label{sec:inf_times}
	
\begin{figure*}[t]
		\centering
		
		\begin{subfigure}[t]{0.48\textwidth}
			\centering
			\includegraphics[width=0.9\linewidth]{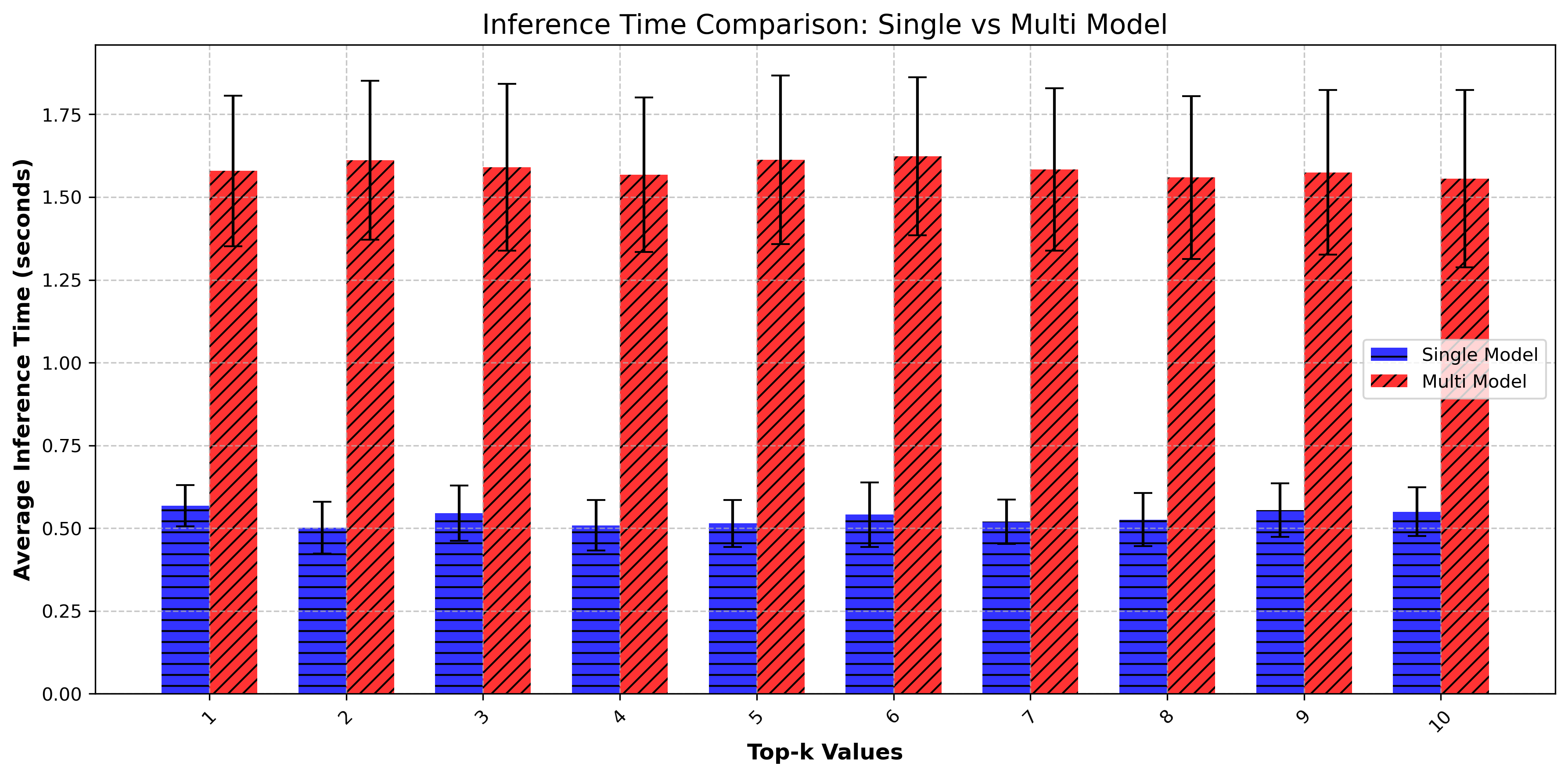}
			\caption{Inference times on Machine \texttt{M1} under the FHE mode.}
			\label{fig:inf_timesi9}
		\end{subfigure}
		\hfill
		\begin{subfigure}[t]{0.48\textwidth}
			\centering
			\includegraphics[width=0.9\linewidth]{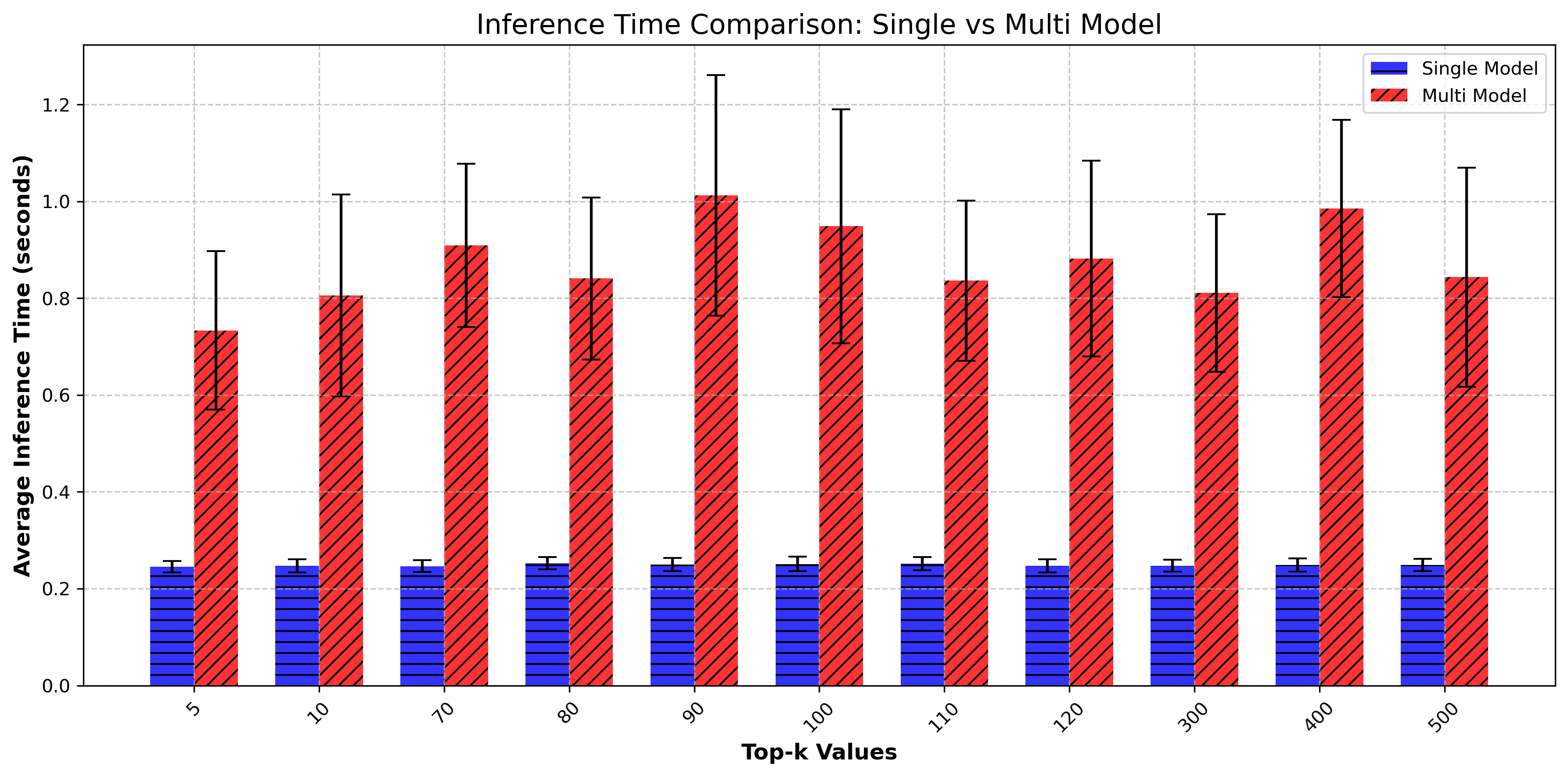}
			\caption{Inference times on Machine \texttt{M3} under the FHE mode.}
			\label{fig:inf_timesAMD}
		\end{subfigure}
		
		\vspace{1em}

		\begin{subfigure}[t]{0.48\textwidth}
			\centering
			\includegraphics[width=0.9\linewidth]{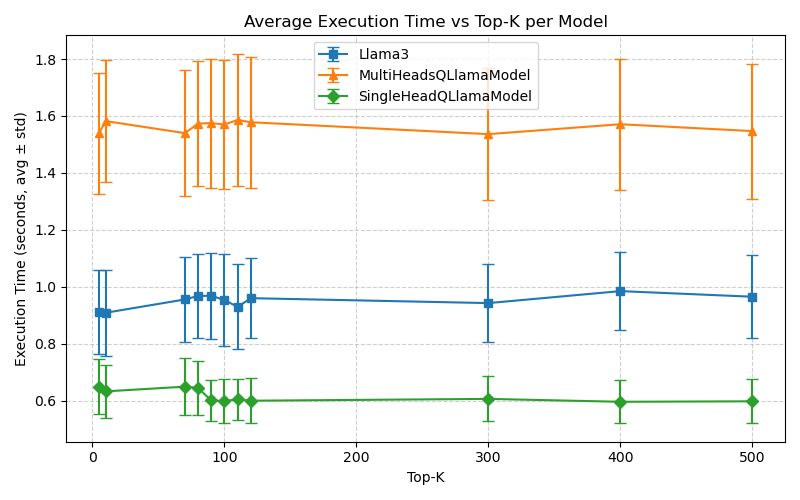}
			\caption{Average text generation times across models on \texttt{M1}.}
			\label{fig:avg_exec_time_M1}
		\end{subfigure}
		\hfill
		\begin{subfigure}[t]{0.48\textwidth}
			\centering
			\includegraphics[width=0.9\linewidth]{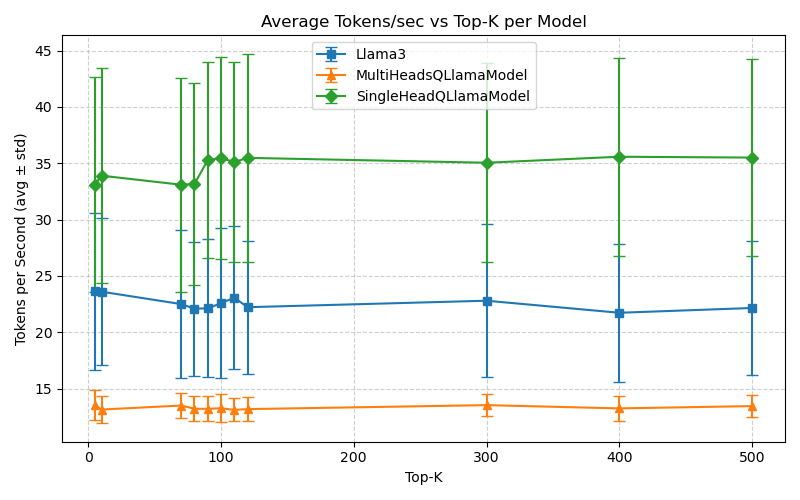}
			\caption{Average tokens generated per second on \texttt{M1}.}
			\label{fig:avg_tokens_per_sec_M1}
		\end{subfigure}
		
		\vspace{1em}
		
		\begin{subfigure}[t]{0.48\textwidth}
			\centering
			\includegraphics[width=0.9\linewidth]{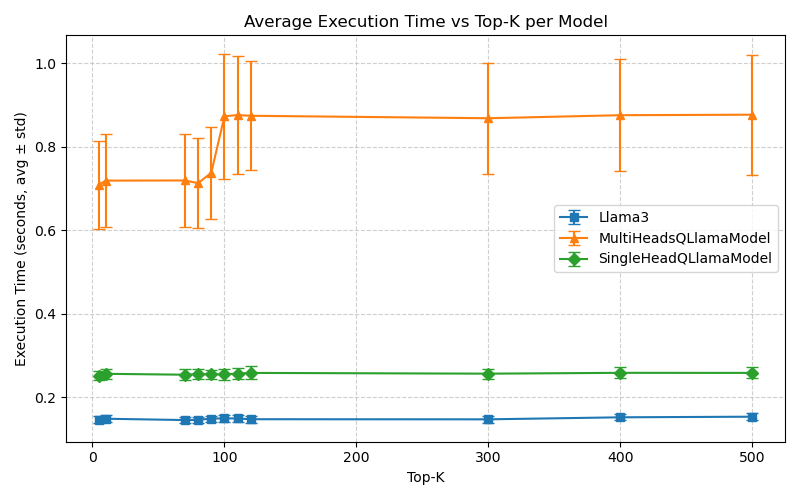}
			\caption{Average text generation times across models on \texttt{M3}.}
			\label{fig:avg_exec_time_M3}
		\end{subfigure}
		\hfill
		\begin{subfigure}[t]{0.48\textwidth}
			\centering
			\includegraphics[width=0.9\linewidth]{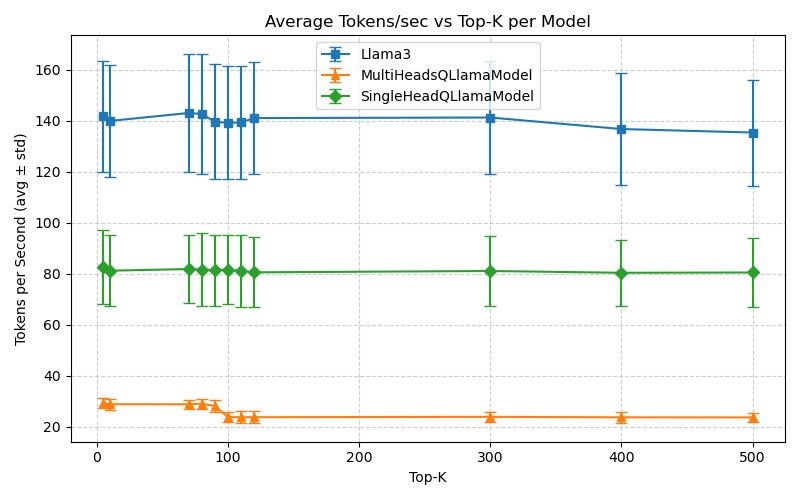}
			\caption{Average tokens generated per second on \texttt{M3}.}
			\label{fig:avg_tokens_per_sec_M3}
		\end{subfigure}

		\caption{Performance comparison of the \textsc{PQC-LLaMA-3} and the plain \textsc{LLaMA-3} models across multiple machines and configurations. Subfigures \ref{fig:inf_timesi9} and \ref{fig:inf_timesAMD} show inference time comparisons on different machines, while [\ref{fig:avg_exec_time_M1}, \ref{fig:avg_tokens_per_sec_M1}] and [\ref{fig:avg_exec_time_M3}, \ref{fig:avg_tokens_per_sec_M3}] pairs present the averaged performance metrics [\texttt{inference times}, \texttt{tokens per second}]  across models.}
		\label{fig:perf_comparison}
	\end{figure*}

	In this appendix, we focus on the best and the worst performing machines \texttt{M1} and \texttt{M3}, respectively, from the described evaluation setup in \ref{sec:exp}. The inference times of the encrypted models \texttt{SingleHeadQLlamaModel} and \texttt{MultiHeadsQLlamaModel} on both machines \texttt{M1} and \texttt{M3} (Figures \ref{fig:inf_timesi9} and \ref{fig:inf_timesAMD}) show the performance lead of the single head encrypted model over the multi-heads one. This behaviour is expected since the computational overhead is greater in the \texttt{MultiHeadsQLlamaModel} incurring many more ciphertext multiplications and relinearization (bootstrapping) operations to regularize the noise levels. Moreover, models' uniformity or fluctuation in the execution times is linked to the main (DRAM) and cache memory accesses and their read-write patterns, which depends on their size and the growing memory requirements by the models.   
	
	Regarding the Tokens/Sec rates illustrated in Figures~\ref{fig:avg_tokens_per_sec_M1} and \ref{fig:avg_tokens_per_sec_M3}, the overall observation and analysis show a link between the tokens/sec rates and the processor's compute power and the memory cache's capacity and architecture. 
	The performance comparison in Figure \ref{fig:avg_tokens_per_sec_M1} shows a clear differentiation in computational efficiency across the three models. The \texttt{SingleHeadQLlamaModel} achieves the highest throughput, averaging $\approx 35$-$37$ tokens/s, outperforming the plain \textsc{LLaMA-3} model, which sustains $\approx 22$-$24$ tokens/s, and significantly exceeding the \texttt{MultiHeadsQLlamaModel}, which achieves $\approx 12$-$13$ tokens/s. This improvement represents an approximate $1.5\times$ increase over the plain model and nearly $3\times$ over the multi-head version, highlighting the efficiency of the single-head FHE-quantized inference configuration.
	
	On the other hand, the models performances are slightly different, mainly the switch between \texttt{SingleHeadQLlamaModel} and \texttt{LLaMA-3} in the order of performance superiority. The general behaviour for machine \texttt{M3} exhibits an overall higher tokens/sec rates compared to \texttt{M1}, due to the latter's higher CPU frequency ($3.2$GHz compared to $2.65$GHz for base frequency, and $6$GHz compared to $3$GHz for turbo clock). The plain \textsc{LLaMA-3} model achieves $142$ tokens/sec which is $1.7\times$ higher than \texttt{SingleHeadQLlamaModel} ($81$ token/sec) and $4.7\times$ higher than \texttt{MultiHeadsQLlamaModel} ($30$ tokens/sec). 
	
	This behaviour described in Figures \ref{fig:avg_exec_time_M1},\ref{fig:avg_exec_time_M3}, \ref{fig:avg_tokens_per_sec_M1},\ref{fig:avg_tokens_per_sec_M3} taken on machines \texttt{M1} and \texttt{M3}, suggests that the reduction in attention head dimensionality and quantized arithmetic under FHE can lead to faster effective throughput, particularly when cache locality and memory access patterns are optimized. Conversely, the multi-head FHE model exhibits a compounded latency due to increased ciphertext management and parallel homomorphic operations across multiple heads, which increase synchronization and memory overhead.
	Furthermore, the cache memory architecture in both machines hints an important potential cause for the superiority of \texttt{SingleHeadQLlamaModel} over \texttt{Llama3} shown in Figure~\ref{fig:avg_tokens_per_sec_M1}. The \texttt{L1} cache size in \texttt{M1} is slightly higher than \texttt{M3}'s which might be a decisive factor when ciphertexts, activations, and gradients are moved to the cache. The larger size of the cache's first level may accommodate ciphertext matrices for the single head model (in \texttt{M1}) while it might be also moved to the next cache level if the $L1$'s is narrower (as in \texttt{M3}). This memory capacity difference might render the single head encrypted model more efficient than the plain one, only due to the cache memory's capacity.

	\section{Cache Memory Readings}\label{sec:cache}

	\begin{figure*}[!ht] 
		\centering
		\begin{subfigure}[ht]{0.48\textwidth}
			\centering
			\includegraphics[width=0.78\linewidth]{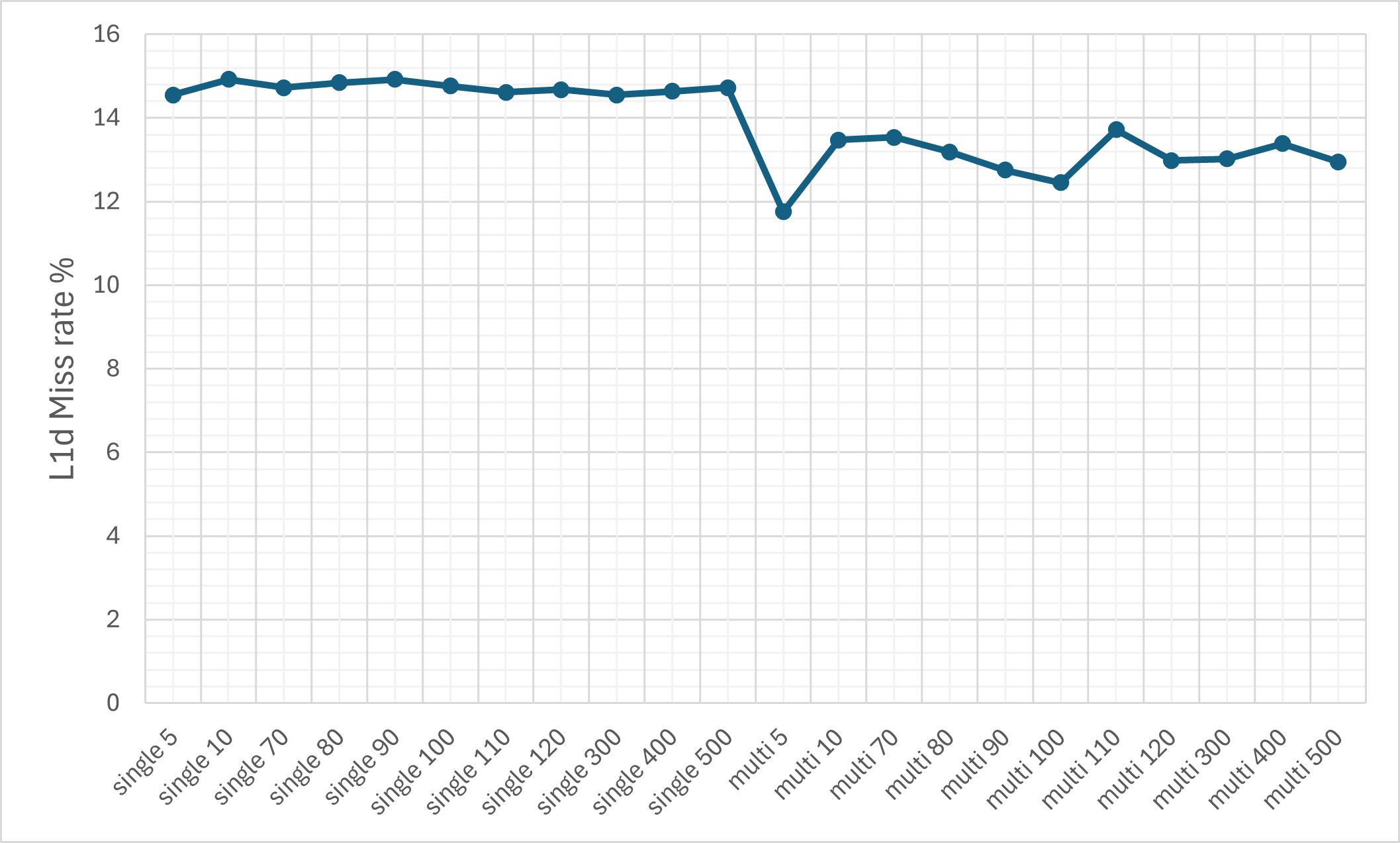}
			\caption{The L1d cache miss rate on the \texttt{M1} machine}
			\label{fig:sub1_m1}
		\end{subfigure}
		\hfill
		\begin{subfigure}[ht]{0.48\textwidth}
			\centering
			\includegraphics[width=0.78\linewidth]{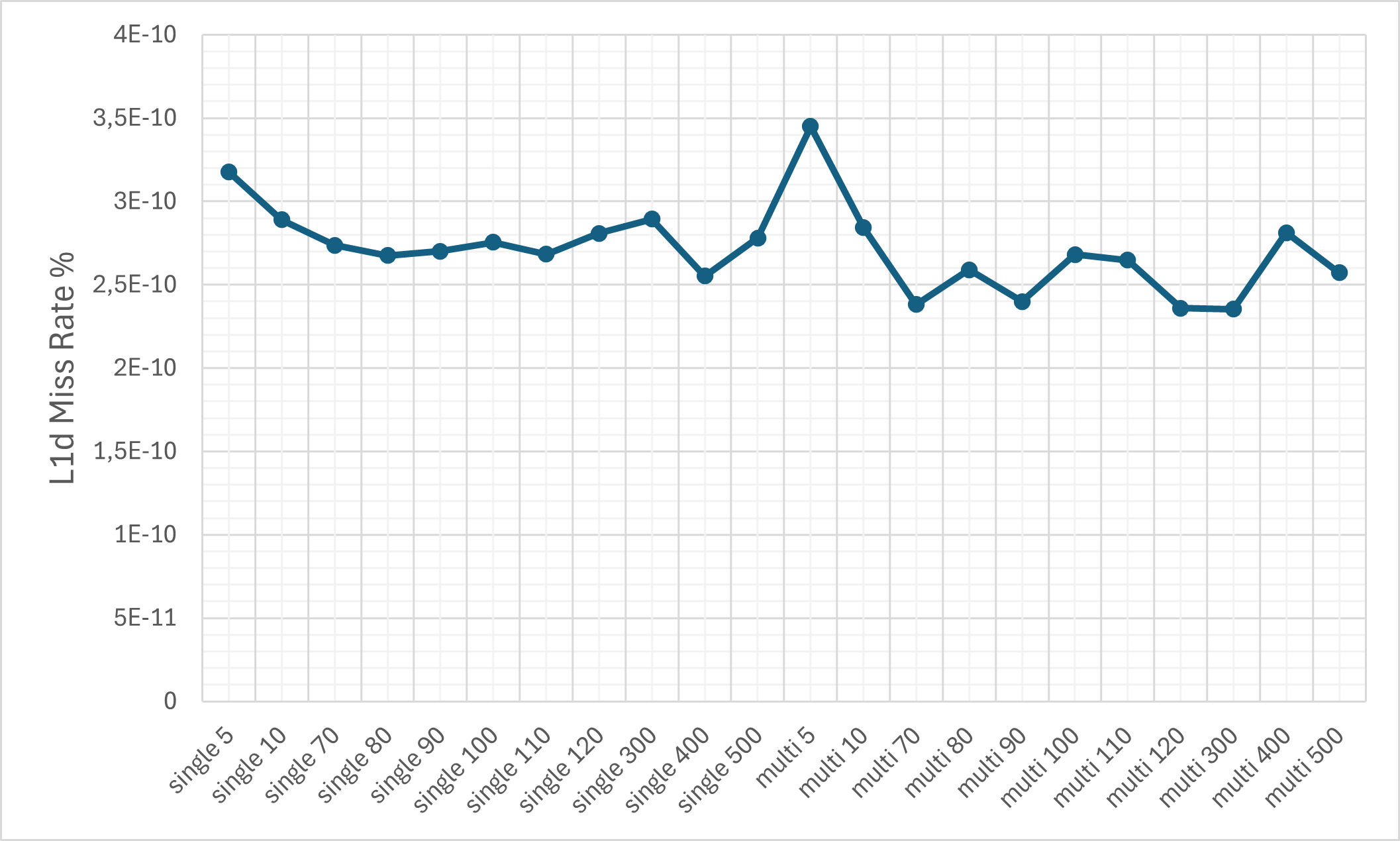}
			\caption{The L1d cache miss rate on the \texttt{M3} machine}
			\label{fig:sub1}
		\end{subfigure}

		\vspace{1em} 
		
		\begin{subfigure}[ht]{0.48\textwidth}
			\centering
			\includegraphics[width=0.78\linewidth]{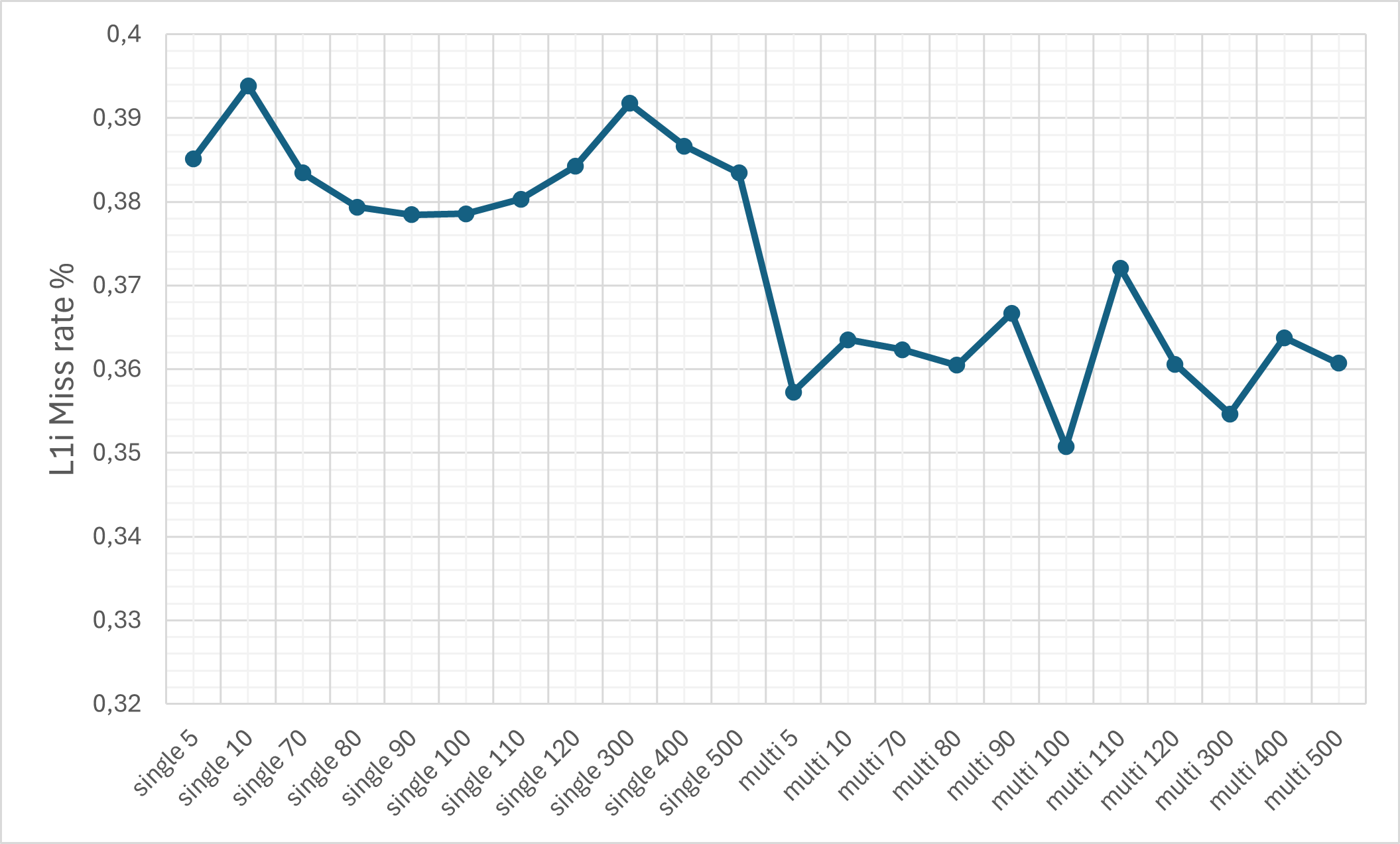}
			\caption{The L1i cache miss rate on the \texttt{M1} machine}
			\label{fig:sub2_m1}
		\end{subfigure}    
		\hfill
		\begin{subfigure}[ht]{0.48\textwidth}
			\centering
			\includegraphics[width=0.78\linewidth]{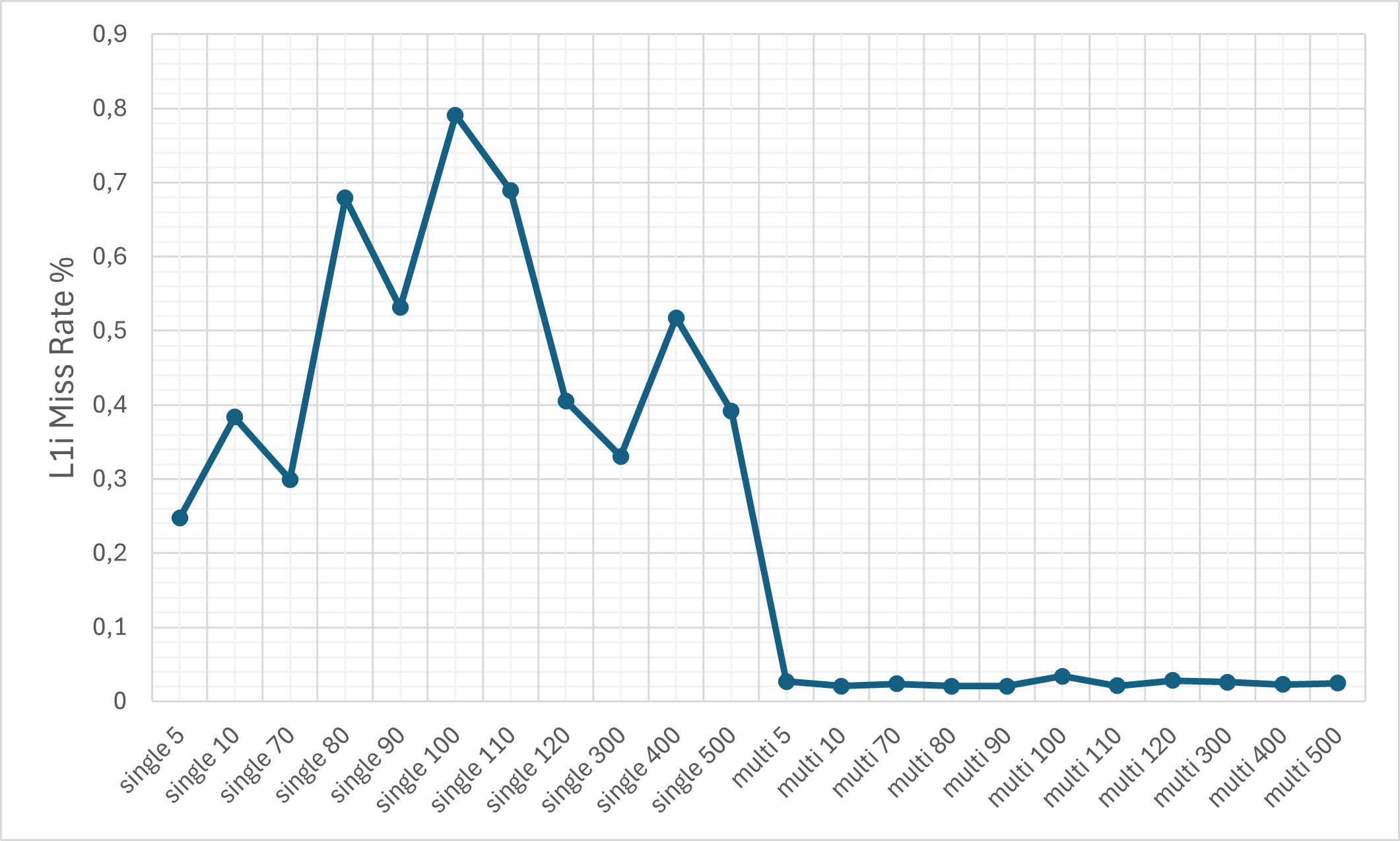}
			\caption{The L1i cache miss rate on the \texttt{M3} machine}
			\label{fig:sub2}
		\end{subfigure}

		\vspace{1em} 
		\begin{subfigure}[ht]{0.48\textwidth}
			\centering
			\includegraphics[width=0.78\linewidth]{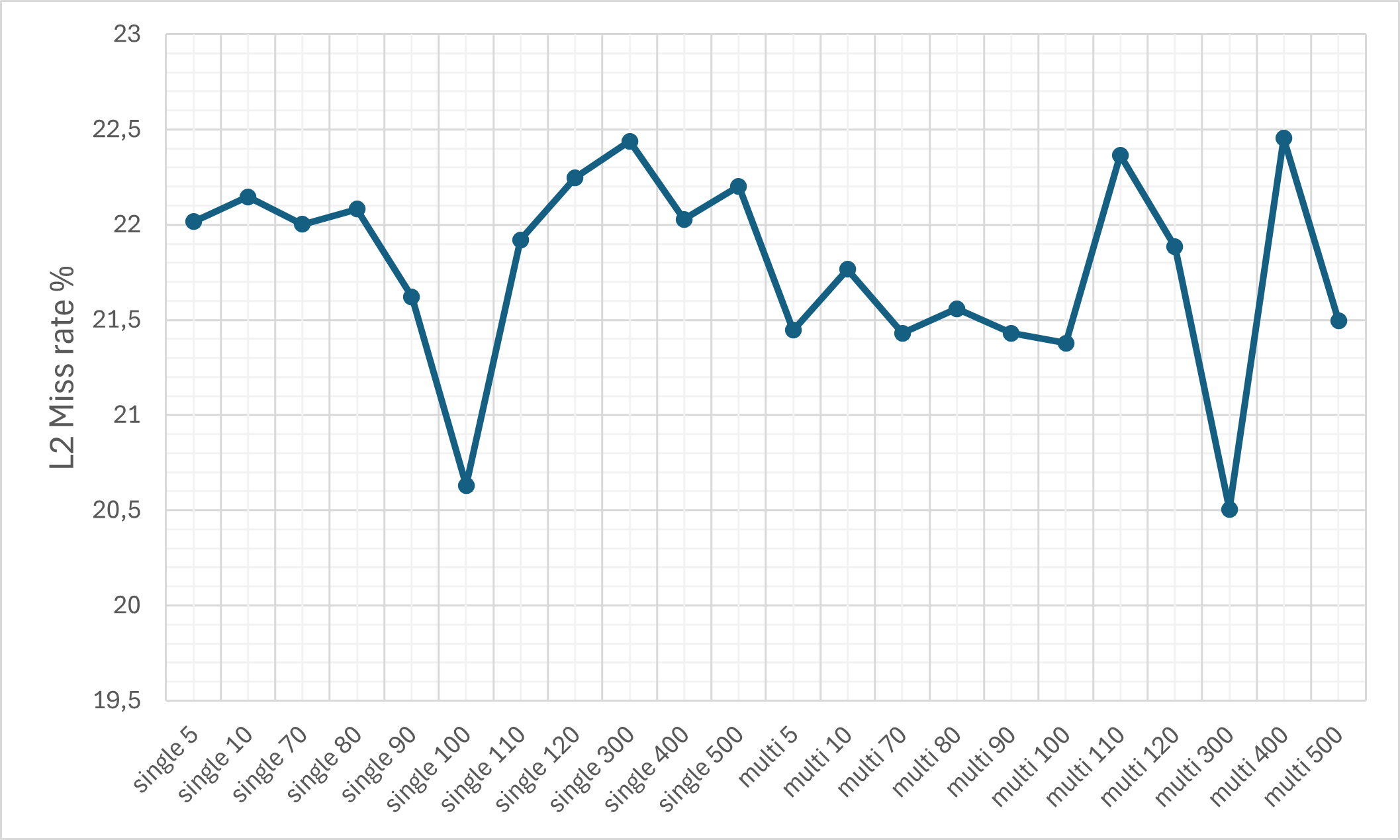}
			\caption{The L2 cache miss rate on the \texttt{M1} machine}
			\label{fig:sub3_m1}
		\end{subfigure}
		\hfill
		\begin{subfigure}[ht]{0.48\textwidth}
			\centering
			\includegraphics[width=0.78\linewidth]{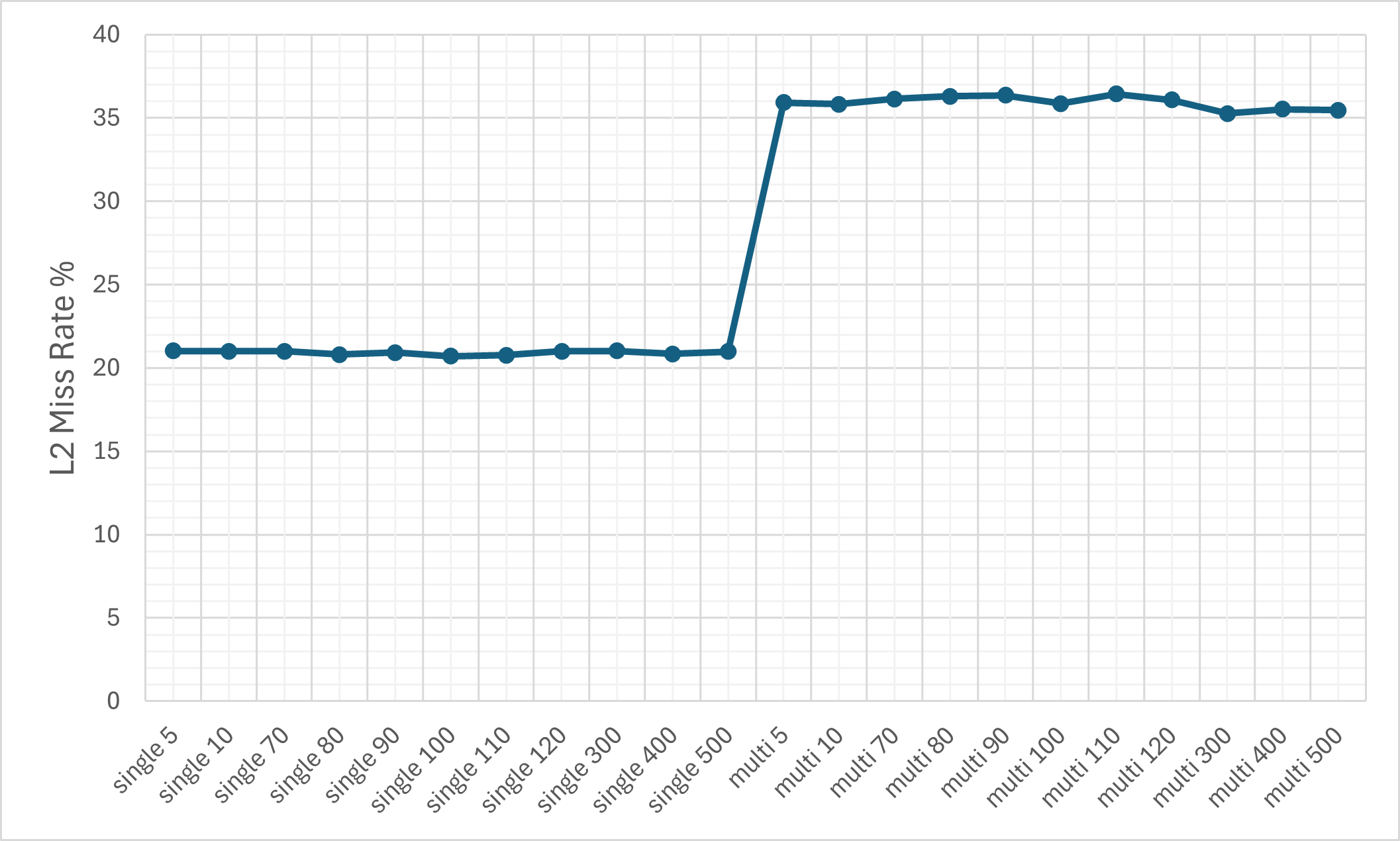}
			\caption{The L2 cache miss rate on the \texttt{M3} machine}
			\label{fig:sub3}
		\end{subfigure}

		\vspace{1em} 
		
		\begin{subfigure}[ht]{0.48\textwidth}
			\centering
			\includegraphics[width=0.78\linewidth]{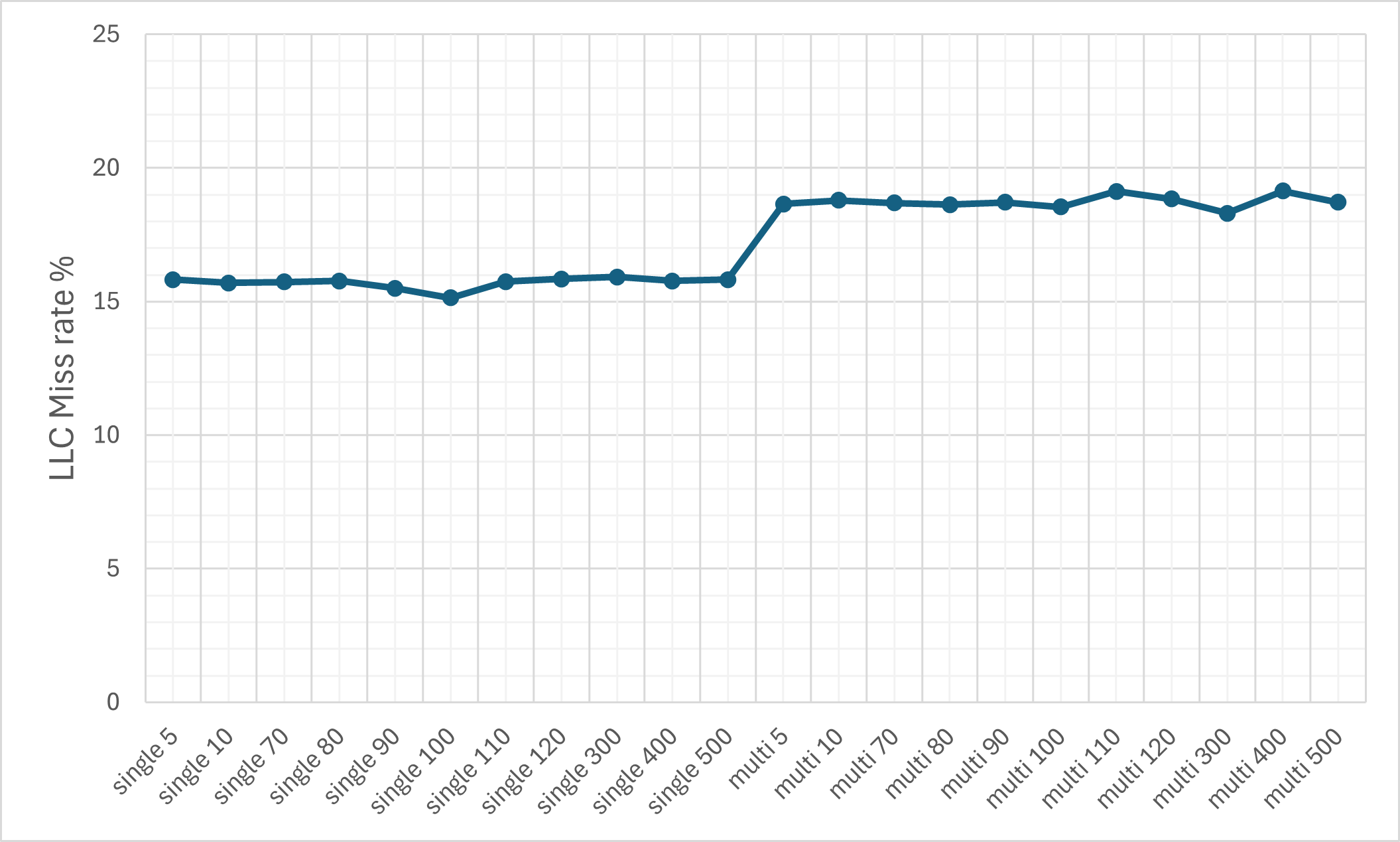}
			\caption{The LLC miss rate on the \texttt{M1} machine}
			\label{fig:sub4_m1}
		\end{subfigure}
		\hfill
		\begin{subfigure}[ht]{0.48\textwidth}
			\centering
			\includegraphics[width=0.78\linewidth]{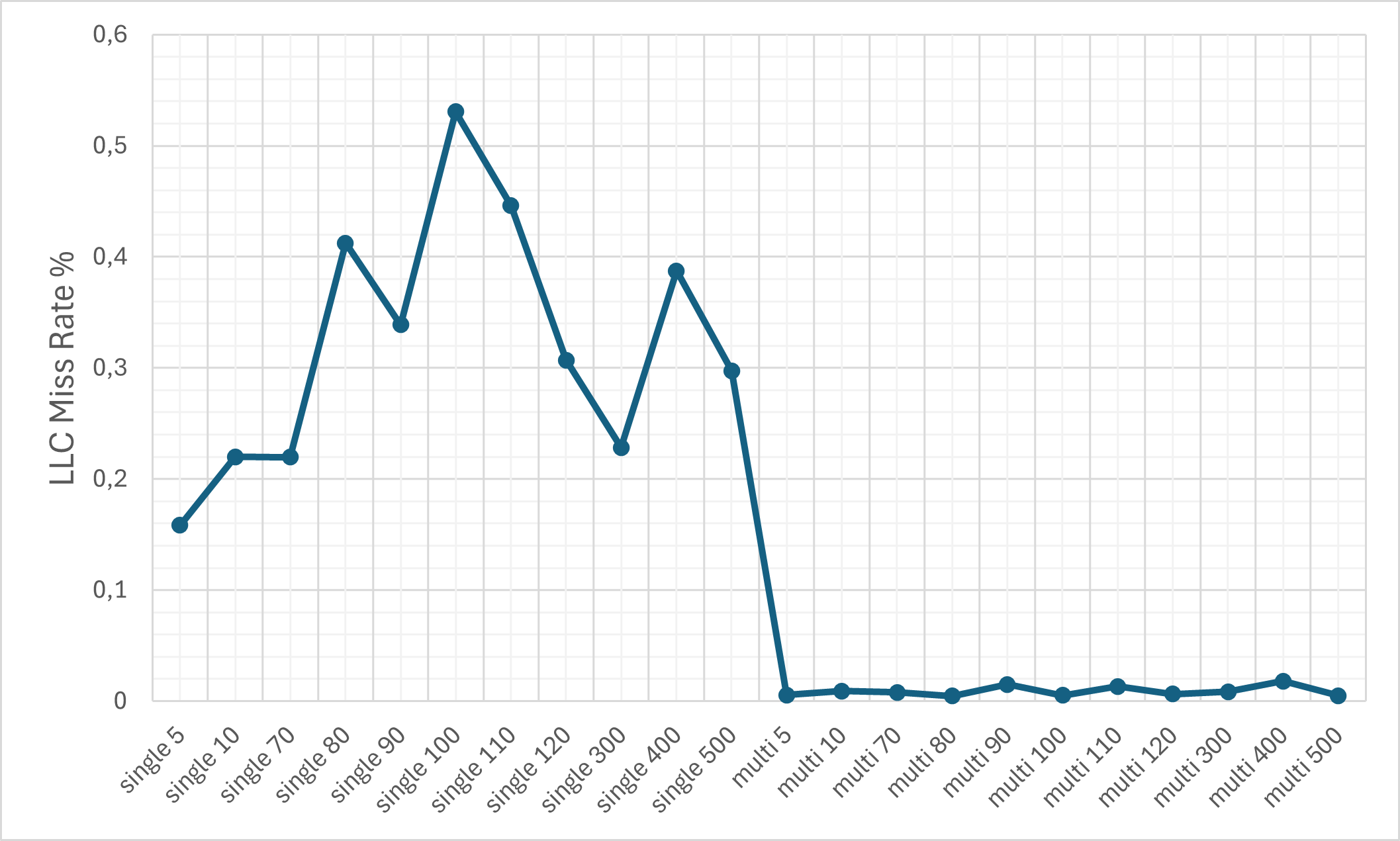}
			\caption{The Last Level Cache (LLC) miss rate on the \texttt{M3} machine}
			\label{fig:sub4}
		\end{subfigure}

		\caption{The cache memory's miss rate for both encrypted \texttt{SingleHeadQLlamaModel} and \texttt{MultiHeadsQLlamaModel} models. The experiments incorporated top-$k$ values varying from short context to a long context text generation. }
		\label{fig:all}
	\end{figure*}

	\begin{figure*}[!ht] 
		\centering
		\begin{subfigure}[ht]{0.48\textwidth}
			\centering
			\includegraphics[width=0.78\linewidth]{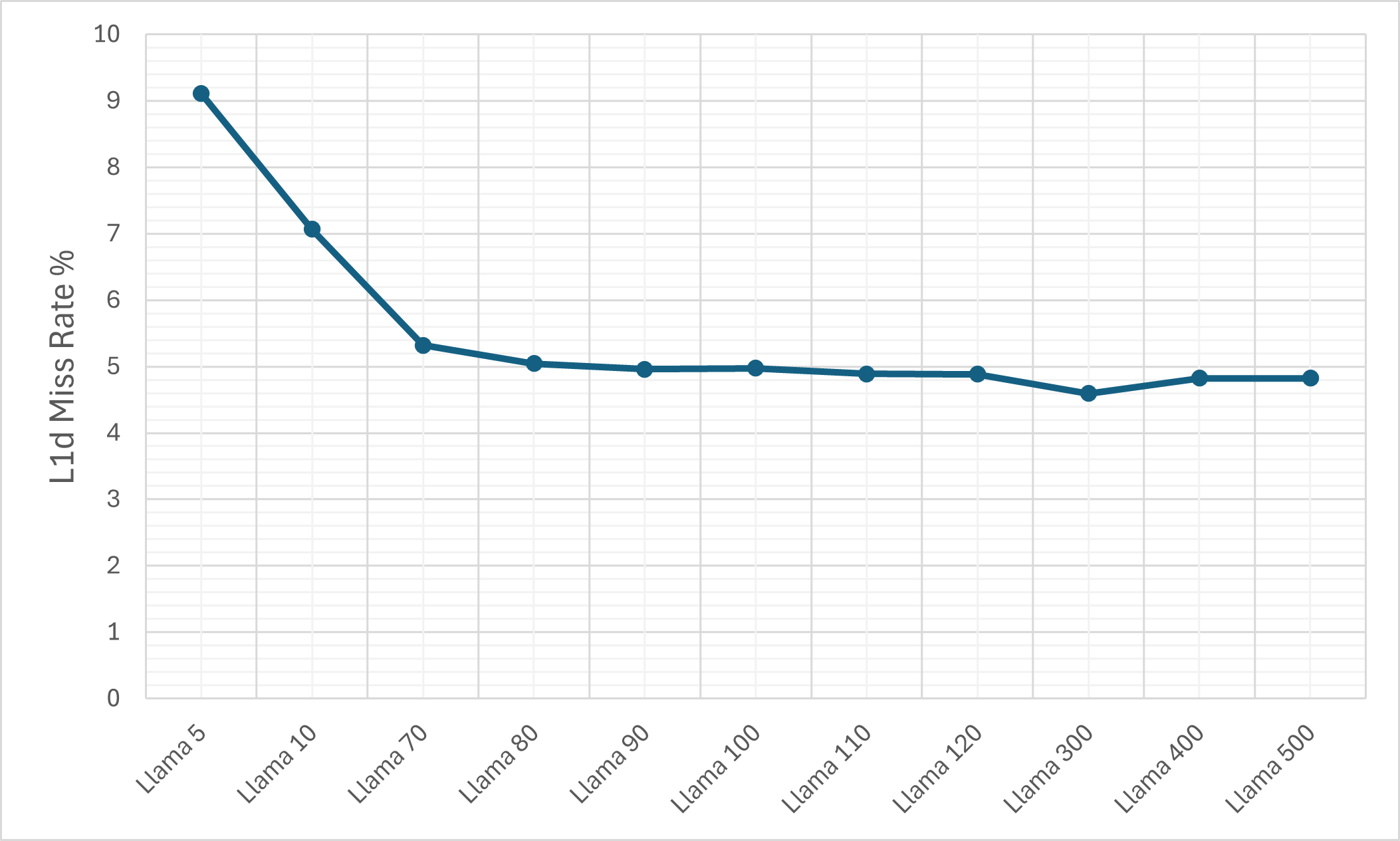}
			\caption{The L1d cache miss rate on the \texttt{M1} machine for the plain \textsc{LLaMA-3} model.}
			\label{fig:sub1_m1_plain_l1d}
		\end{subfigure}
		\hfill
		\begin{subfigure}[ht]{0.48\textwidth}
			\centering
			\includegraphics[width=0.78\linewidth]{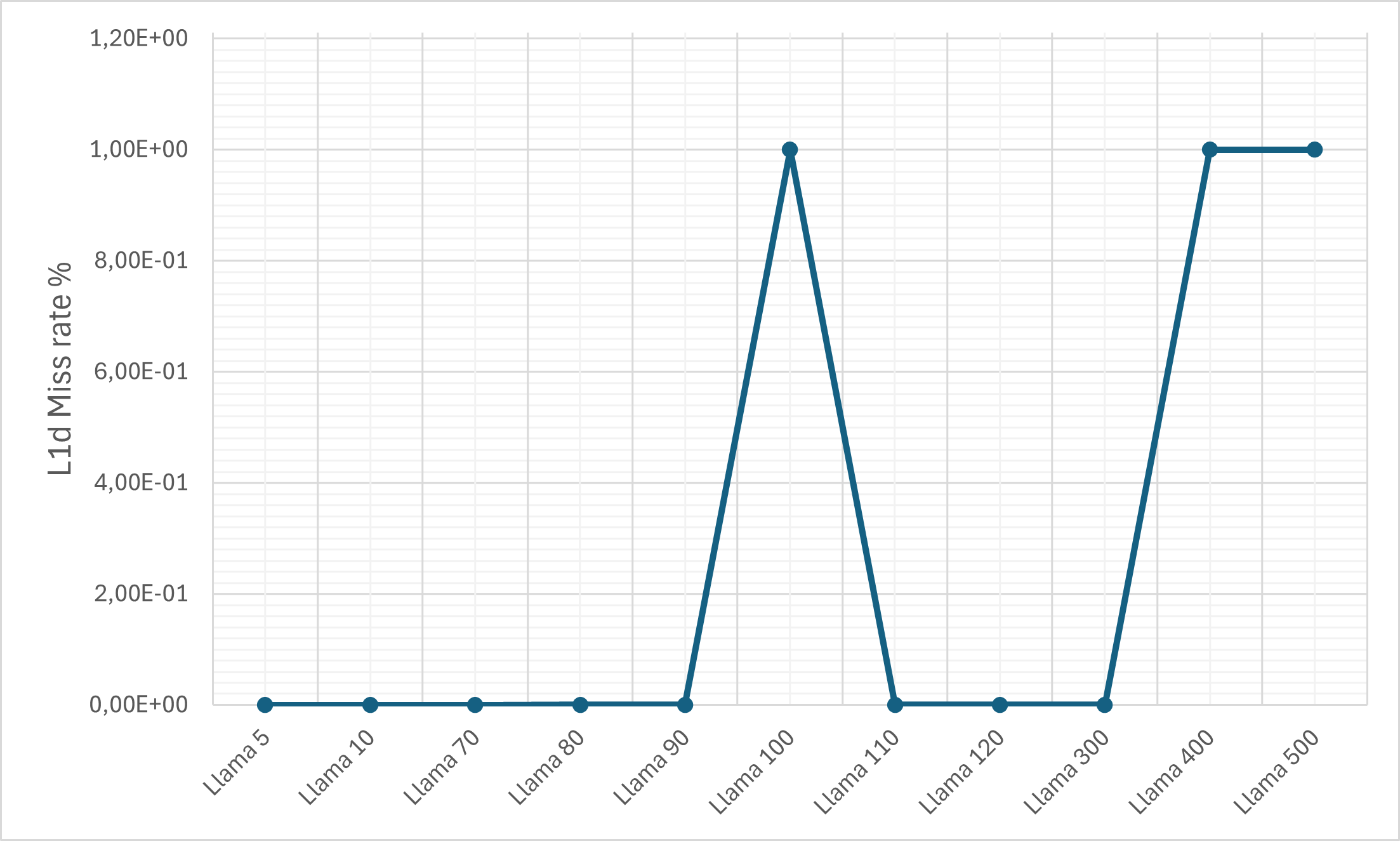}
			\caption{The L1d cache miss rate on the \texttt{M3} machine for the plain \textsc{LLaMA-3} model}
			\label{fig:sub1_m3_plain}
		\end{subfigure}
		
		\vspace{1em}

		\begin{subfigure}[ht]{0.48\textwidth}
			\centering
			\includegraphics[width=0.78\linewidth]{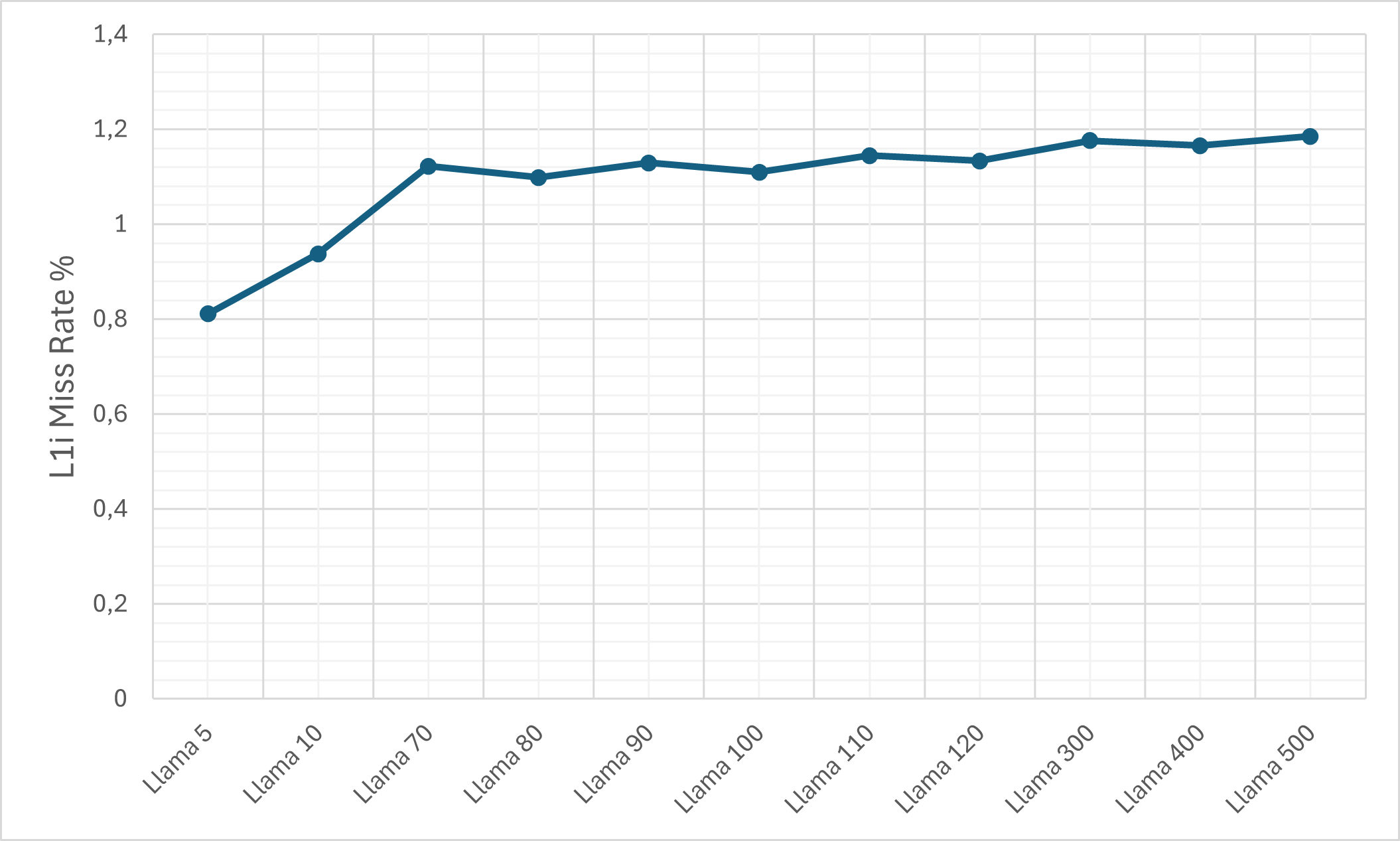}
			\caption{The L1i cache miss rate on the \texttt{M1} machine for the plain \textsc{LLaMA-3} model.}
			\label{fig:sub1_m1_plain_l1i}
		\end{subfigure}
		\hfill
		\begin{subfigure}[ht]{0.48\textwidth}
			\centering
			\includegraphics[width=0.78\linewidth]{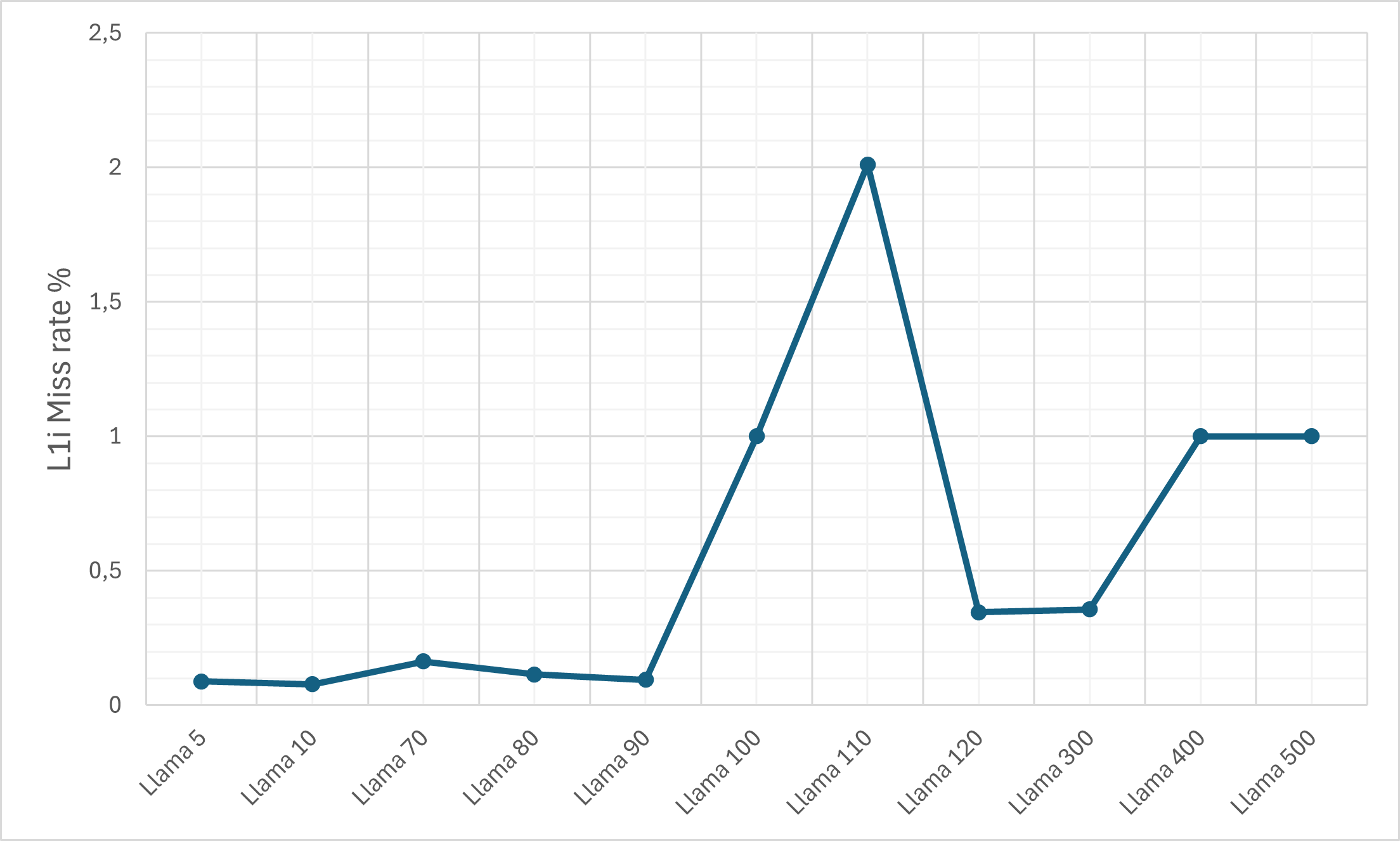}
			\caption{The L1i cache miss rate on the \texttt{M3} machine for the plain \textsc{LLaMA-3} model}
			\label{fig:sub2_m3_plain}
		\end{subfigure}
		
		\vspace{1em}
		
		\begin{subfigure}[ht]{0.48\textwidth}
			\centering
			\includegraphics[width=0.78\linewidth]{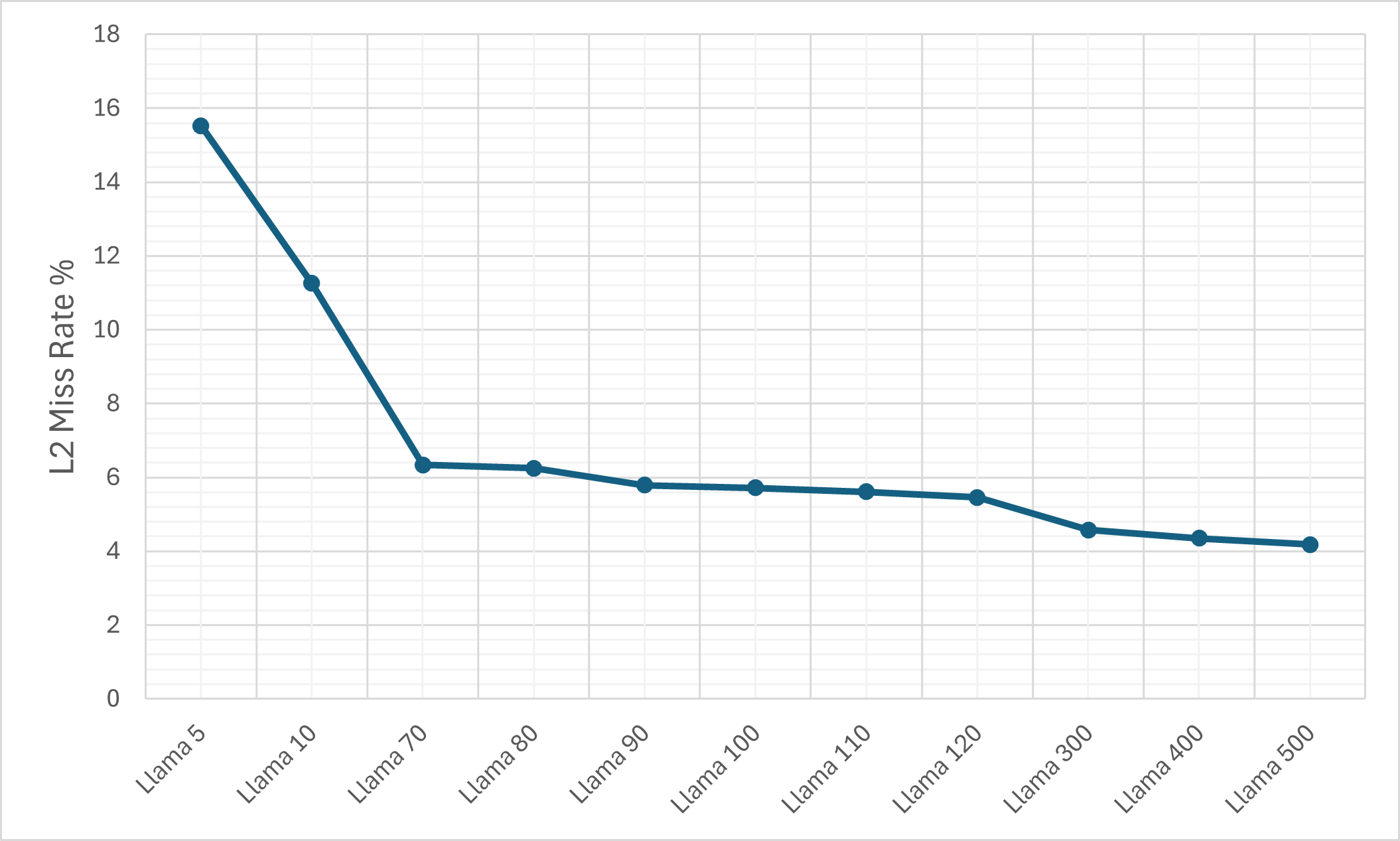}
			\caption{The L2 cache miss rate on the \texttt{M1} machine for the plain \textsc{LLaMA-3} model}
			\label{fig:sub1_m1_plain_l2}
		\end{subfigure}
		\hfill
		\begin{subfigure}[ht]{0.48\textwidth}
			\centering
			\includegraphics[width=0.78\linewidth]{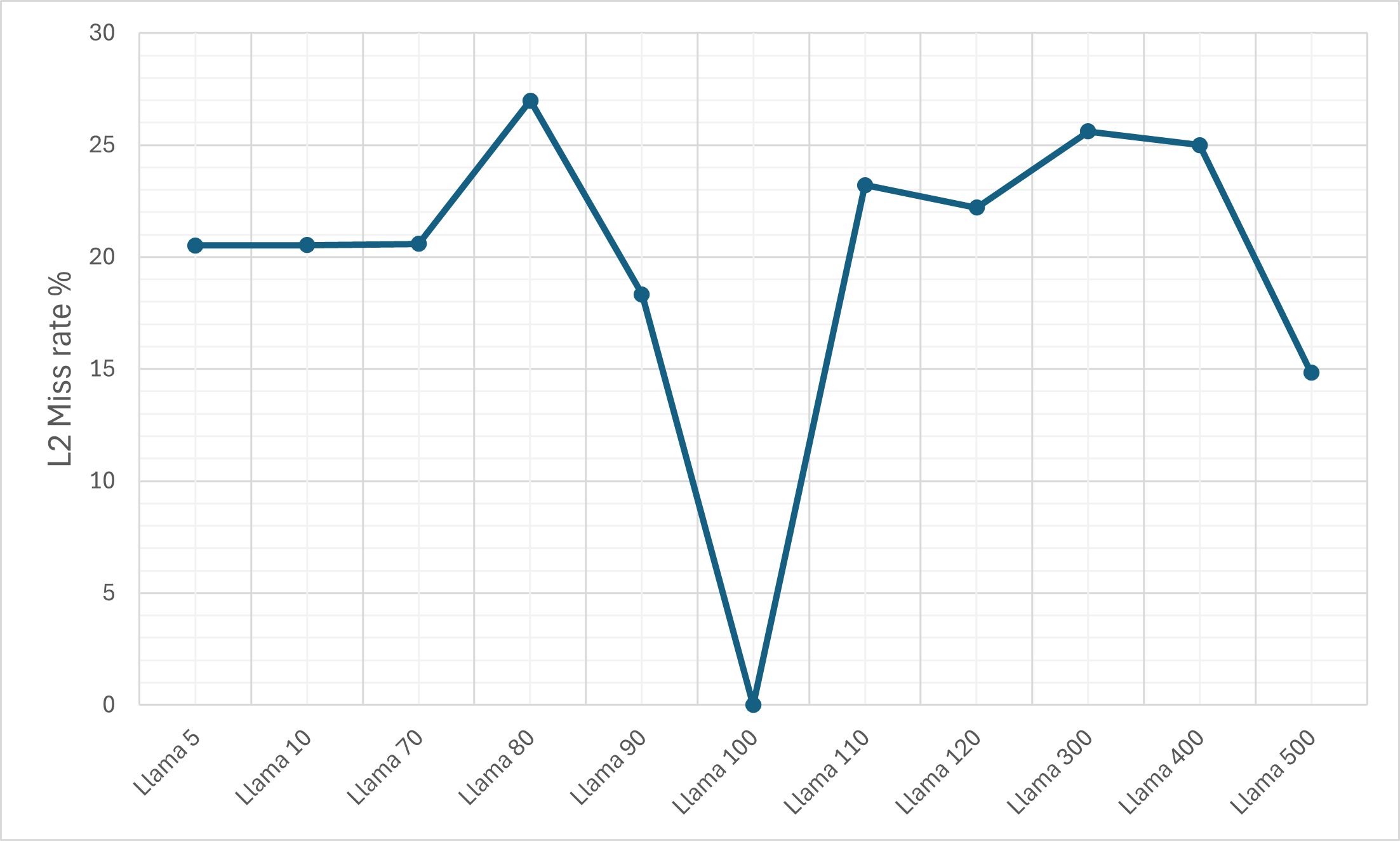}
			\caption{The L2 cache miss rate on the \texttt{M3} machine for the plain \textsc{LLaMA-3} model}
			\label{fig:sub3_m3_plain}
		\end{subfigure}
		
		\vspace{1em}
		
		\begin{subfigure}[ht]{0.48\textwidth}
			\centering
			\includegraphics[width=0.78\linewidth]{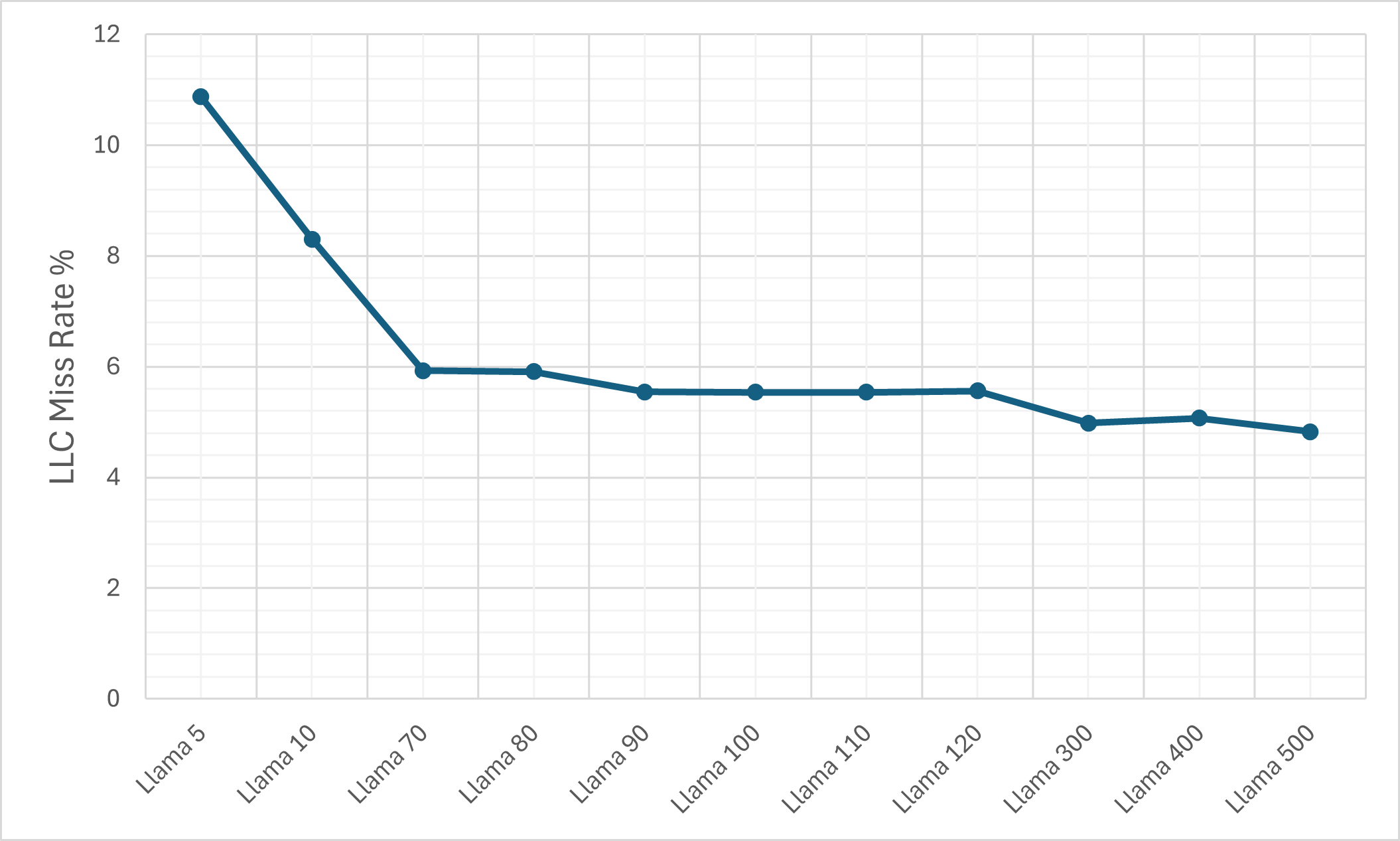}
			\caption{The LLC cache miss rate on the \texttt{M1} machine for the plain \textsc{LLaMA-3} model}
			\label{fig:sub1_m1_plain_llc}
		\end{subfigure}
		\hfill
		\begin{subfigure}[ht]{0.48\textwidth}
			\centering
			\includegraphics[width=0.78\linewidth]{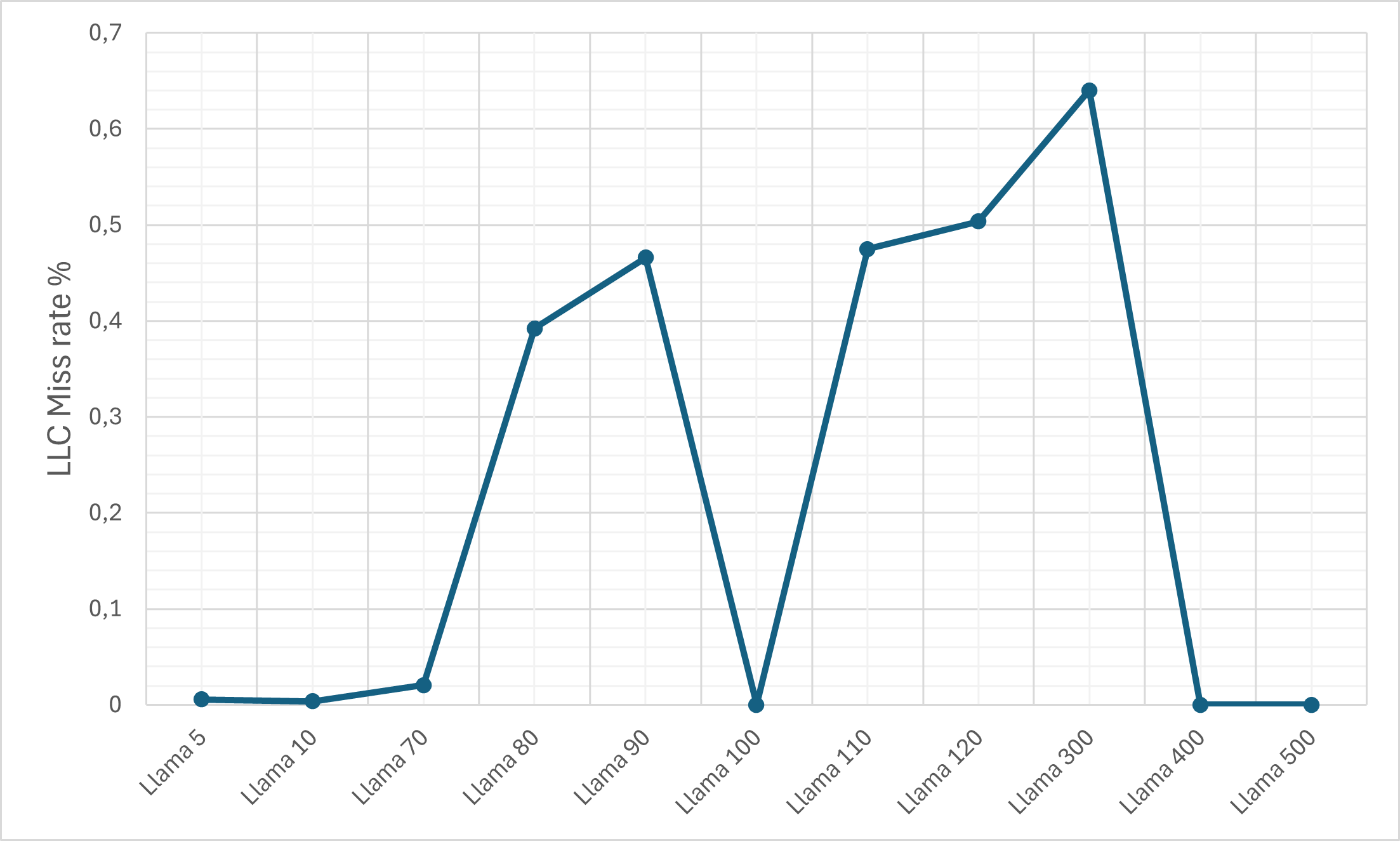}
			\caption{The LLC cache miss rate on the \texttt{M3} machine for the plain \textsc{LLaMA-3} model}
			\label{fig:sub4_m3_plain}
		\end{subfigure}
		
		\caption{The cache memory's miss rate for the plain unencrypted \textsc{LLaMA-3} model. The experiments incorporated top-$k$ values varying from short context to a long context text generation. }
		\label{fig:plain_cache}
	\end{figure*}

	\subsection{Cache Behaviour on \textsc{M1}}
	The memory access patterns on the machine \texttt{M1} is different than \texttt{M3}, the general access pattern shows general steady cache accesses with more dependency on the \texttt{L1} and \texttt{L2} caches and less \texttt{L3/LCC} accesses by all \textsc{PQC-LLaMA-3} models (\texttt{single} and \texttt{multi}). More granular, the \texttt{MultiHeadsQLlamaModel} is more efficient than the \texttt{SingleHeadQLlamaModel} regarding the \texttt{L1} and \texttt{L2} cache memory, reaching the following approximate rates; $13\%$ for $15\%$ (Figure \ref{fig:sub1_m1}), $0.36\%$ for $0.385\%$ (Figure \ref{fig:sub2_m1}), and $21.6\%$ for $22.2\%$ (Figure \ref{fig:sub3_m1}), respectively, when comparing these two models. However, this behaviour is reversed while inspecting the \texttt{L3} cache miss rates, the \texttt{MultiHeadsQLlamaModel} slightly degrades in its performance reaching $\approx$ $19\%$ (Figure \ref{fig:sub4_m1}).      
	
	Taken together, these measurements paint a coherent picture of the memory hierarchy behaviour for FHE-quantized inference with \textit{KV-caching}:
	
	\begin{itemize}
		\item Strong \texttt{L1} locality (data and instructions). The very low \texttt{L1d} and \texttt{L1i} miss rates imply that the hot inner loops (FHE kernels, attention scoring, and per-token postprocessing) largely operate on data that is quickly reused and remains resident in \texttt{L1}. This outcome is plausible in a KV-cached inference pipeline where most per-token operations reuse small vectors and matrix fragments that are repeatedly accessed during attention computations.
		\item \texttt{L2} as the pressure point. The elevated \texttt{L2} miss rates point to the \texttt{L2} cache ($24$ MiB) being the principal capacity/associativity boundary for the workload. Two complementary mechanisms likely explain this:
		\begin{enumerate}
			\item Working-set size and streaming patterns. FHE-quantized tensors (ciphertext blocks, quantized weight shards) and per-token state can be large. Although the hot inner working set fits in \texttt{L1}, the aggregate data for many tokens and multi-head attention (multiple heads producing/consuming KV pairs) can exceed \texttt{L2} residency. This leads to \texttt{L1} misses that find \texttt{L2} only intermittently.
			
			\item Parallelism and interleaved streams (multi-head). The multi-head model can create more concurrent memory streams (per-head buffers, parallel attention kernels). Even when total \texttt{LLC} residency is sufficient, the distribution of those streams can increase \texttt{L2} conflict misses (associativity and replacement interactions) and thus raise \texttt{L2} miss fractions.
		\end{enumerate}
		\item \texttt{LLC} largely mitigates DRAM traffic. The large $256$ MiB \texttt{LLC} provides a high-capacity region for sharing and reusing KV data across tokens and heads. The moderate \texttt{LLC} miss rates indicate that although a non-negligible fraction of accesses escalates beyond \texttt{L3}, a majority of \texttt{L2} misses are resolved in \texttt{LLC}. This behaviour explains the modest observed variability of end-to-end latency across top-k: as long as KV caches and frequently reused quantized arrays remain resident in \texttt{L3}, increasing top-k does not substantially increase DRAM traffic or invocation latency.
		
	\end{itemize}

	\subsection{Cache Behaviour on \textsc{M3}}
	
	The Figures \ref{fig:sub1}, \ref{fig:sub2}, \ref{fig:sub3}, \ref{fig:sub4} show the recorded cache memory miss rates in all its levels, \texttt{L1}, \texttt{L2}, and \texttt{L3}/\texttt{LLC}. The general observation for both single head and multi heads models is cache memory dependency and usage efficiency on all levels (except for \texttt{L2} as a special case), suggesting different spatial and temporal locality behaviours influenced by the encryption overhead and the model’s internal parallelism.
	
	The \texttt{L1d} cache's miss rate is extremely low across all top-$k$ values for each model ($\approx 3.5\times10^{-12}$), showcasing that the models weight tensors and some activation blocks reside in the first cache memory's level. Nevertheless, there are some micro observations to be made. As shown in Figure \ref{fig:sub1}, the highest miss rates belong to \texttt{single 5} and \texttt{multi 5} trials, which is due to the cache warmup condition. The cache needs to be populated before start storing and fetching data for the running program, in our scenario, the \texttt{L1d} needs to be populated with the model's weights so it would retrieve it faster for next calls.
	Furthermore, the \texttt{MultiHeadsQLlamaModel} has a slightly better miss rate than the \texttt{SingleHeadQLlamaModel}, which can be noticed at the data point starting from \texttt{multi 10} (we excluded \texttt{multi 5} due to the warmup step).
	
	For the L1 instruction cache (\texttt{L1i}), the single-head model shows moderate variation across top-$k$ values, particularly a noticeable increase in miss rate from \texttt{single 70} to \texttt{single 100}. This is attributed to branch and kernel-switching overhead caused by token-dependent control flow during longer-context inference. The multi-head model maintains much lower \texttt{L1i} miss rates ($\leq$ $0.1\%$), indicating better instruction-level locality. This improvement likely arises from instruction reuse across parallel attention heads, where identical kernel sequences are reused, reducing instruction fetch misses.
	
	The situation is different with the second level of the cache \texttt{L2} (Figure \ref{fig:sub3}). For both models, the miss rate is relatively high (when compared with the other levels), with higher rates for the multi heads model (with averages around $21\%$ for the single-head and $36\%$ for the multi-head model). It appears that the models are less dependent on \texttt{L2} with more emphasis on the other cache levels. For FHE workloads, such a behaviour may be linked to the intermediate ciphertext buffers, which are relatively large and often would not fit the \texttt{L2} cache due to frequent eviction when the encrypted attention blocks exceed the \texttt{L2} capacity. The multi-head model, as noticed in Figure \ref{fig:sub3} starting from \texttt{multi 5}, arouse this effect because parallel encrypted heads introduce multiple concurrent and large memory streams, increasing \texttt{L2} contention and decreasing its effective reuse distance.
	
	The cache access patterns for \texttt{L1i} (the first cache level for instructions) and \texttt{LLC} (or \texttt{L3} cache) are quite similar (see Figures \ref{fig:sub2} and \ref{fig:sub4}). The single head model shows a fluctuating cache access but still efficient ($0.15$–$0.8\%$, range including both cache levels \texttt{L1i} and \texttt{LLC}). Moreover, it experiences degradation in the cache access when moving from short context to a longer context text generation (from $10$ to $500$ new tokens) on both cache levels, \texttt{L1i} and \texttt{LLC}. This can be linked to the sudden change from short to longer new token sequences where the in-between tokens need to be computed first (they were not computed before) then stored in the cache before fetching them. This cache access behaviour would be different if intermediate trials were present between \texttt{single 10} and \texttt{single 70}. On the other hand, the multi heads model reaches a perfect \texttt{LLC} hit rate while performing steadily across all top-$k$ values, reflecting effective sharing of precomputed ciphertexts and quantized tensors across the attention heads. 
	The steady \texttt{LLC} hit rate also helps explain why the average inference times remain nearly constant across all top-$k$ values, once the model’s working set fits within the \texttt{LLC}, additional tokens do not significantly increase cache misses or latency.
	
	Overall, the results indicate that both models exhibit strong \texttt{L1} and \texttt{LLC} locality, but diverge at the \texttt{L2} level due to architectural and the required memory budget imposed by the integration of FHE primitives and functions within the model. The \texttt{MultiHeadsQLlamaModel} demonstrates higher \texttt{L2} miss rates yet achieves superior \texttt{LLC} efficiency, implying that data sharing at the last-level cache compensates for mid-level cache inefficiency. Consequently, the similar inference times observed across top-$k$ values are justified by the stable \texttt{L1d} and \texttt{LLC} hit behaviour, which mitigates the latency impact of variable top-$k$ decoding workloads.

	\subsection{Cache Behaviour with LLaMA-3}
	
	Figure \ref{fig:plain_cache} presents the cache miss rates for the plain (unencrypted) \textsc{LLaMA-3} model across different top-$k$ configurations, alongside the previously measured Fully Homomorphic Encryption quantized single-head and multi-head variants. The plain model demonstrates a consistently low miss rate at all cache levels, reflecting a more efficient memory access pattern and reduced cache pressure. Specifically, the \texttt{L1d} data miss rate for the plain model decreases from $9.1\%$ at top-$k$ = $5$ to approximately $4.8\%$ beyond top-$k$ = $100$ (Figure \ref{fig:sub1_m1_plain_l1d}), representing a $2$-$3\times$ improvement compared to the encrypted single-head ($\approx$ $14$-$15\%$) and multi-head ($\approx$ $12$-$13\%$) counterparts on \texttt{M1} (Figure \ref{fig:sub1_m1}). This reduction suggests that the unencrypted model exhibits better temporal and spatial locality in weight and activation reuse, allowing more effective exploitation of the $1.5$ MiB \texttt{L1d} data cache.
	
	A similar trend is observed at the \texttt{L2} and \texttt{LLC} levels, where the plain model’s miss rates fall within $4$-$15\%$ and $5$-$11\%$, respectively, (see Figures \ref{fig:sub1_m1_plain_l2} and \ref{fig:sub1_m1_plain_llc}) significantly lower than the encrypted models ($20$-$22\%$ for \texttt{L2} and $15$-$19\%$ for \texttt{L3}) on machine \texttt{M1} (Figures \ref{fig:sub3_m1} and \ref{fig:sub4_m1}). The sharper decline in \texttt{L2} and \texttt{L3} miss rates for higher top-$k$ values indicates that larger decoding workloads favor cache reuse under plaintext arithmetic, whereas encrypted execution under FHE incurs additional tensor expansion and ciphertext-related overhead, which leads to less predictable access patterns and more frequent evictions across all cache levels.
	
	Interestingly in Figures \ref{fig:sub2_m1} and \ref{fig:sub1_m1_plain_l1i}, the \texttt{L1i} instruction cache miss rate ($\approx$ $0.8$-$1.2\%$) for the plain model is slightly higher than in the FHE cases ($\approx$ $0.36$-$0.38\%$), which can be attributed to the higher instruction-level diversity and deeper execution path of the native \textsc{LLaMA-3} model. The FHE quantized models, in contrast, exhibit a more uniform execution flow dominated by homomorphic arithmetic kernels with repetitive instruction sequences that fit efficiently within the \texttt{L1} instruction cache.
	
	The cache access patterns for the plain model on machine \texttt{M3} (Figures \ref{fig:sub1_m3_plain}, \ref{fig:sub2_m3_plain}, \ref{fig:sub3_m3_plain}, and \ref{fig:sub4_m3_plain}) have a near binary access mode, either a full cache hit or miss. For instance, when there is a $100\%$ cache miss for \texttt{Llama 100} settings on the first level cache \texttt{L1d}, it is answered by an access from the third level as shown in Figures \ref{fig:sub1_m3_plain}, \ref{fig:sub3_m3_plain}, and \ref{fig:sub4_m3_plain} having a $100\%$ hit rate. The same observation for instances \texttt{Llama 400} and \texttt{Llama 500}. Furthermore, it seems that the plain model depends less on the second cache level \texttt{L2} where the miss rates are between $0\%$ and $27\%$. The cache memory performance for the plain model remains superior to the encrypted models (single and multi heads) on machine \texttt{M3}, however, the difference in performance is too narrow for all cache levels. It is also important to note that the performance of memory access on \texttt{M3} is more compelling than than its counterpart \texttt{M1}.
	
	Overall, These observations demonstrates an efficient use of the cache memory by the plain model and distinguishes it from the other behaviours of the encrypted models, which can be linked to the extra memory and computation overhead imposed by the additional ciphertexts matrices and homomorphic operations in the attention layers.
	
	These results also indicate that while FHE quantization provides cryptographic security and data privacy guarantees, it introduce some cache inefficiency due to ciphertext expansion, larger memory footprints, and reduced data reuse, which is an expected logic. The plain \texttt{LLaMA-3} model, benefiting from compact data representation and regular access patterns, maintains high cache locality across all hierarchy levels. These findings highlight the memory access trade-off inherent to privacy-preserving inference, where encrypted computation leads to measurable degradation in cache efficiency and consequently higher memory latency and energy cost.


\end{appendices}




\end{document}